\documentclass[amsmath,amssymb,aps,prl,twocolumn,superscriptaddress,floatfix,nofootinbib]{revtex4-2}

\usepackage{graphicx}
\usepackage{dcolumn}
\usepackage{bm}
\usepackage{hyperref}
\usepackage{amsmath}
\usepackage{placeins}
\usepackage{color}

\hypersetup{colorlinks, linkcolor = [rgb]{0,0.0,0.75}, citecolor = [rgb]{0,0.0,0.75}, urlcolor = [rgb]{0,0.0,0.75}}

\newcommand{\op}{\mathcal{O}}
\newcommand{\g}[1]{\gamma_{#1}}
\newcommand{\nb}{\phantom{0}}
\newcommand{\wm}{\phantom{-}}

\begin{document}
	
\title{Shallow Bound States and Hints for Broad Resonances with Quark Content \texorpdfstring{$\bm{\bar{b}\bar{c}ud}$}{bcud} \\ in \texorpdfstring{$\bm{B\text{-}\bar{D}}$}{B-D} and \texorpdfstring{$\bm{B^*\text{-}\bar{D}}$}{Bstar-D} Scattering from Lattice QCD}

\author{Constantia Alexandrou}
\affiliation{Department of Physics, University of Cyprus, 20537 Nicosia, Cyprus}
\affiliation{Computation-based Science and Technology Research Center, The Cyprus Institute, 20 Konstantinou Kavafi Street, 2121 Nicosia, Cyprus}

\author{Jacob Finkenrath}
\affiliation{Computation-based Science and Technology Research Center, The Cyprus Institute, 20 Konstantinou Kavafi Street, 2121 Nicosia, Cyprus}
\affiliation{Bergische Universität Wuppertal, Gaußstraße 20, 42119 Wuppertal, Germany}

\author{Theodoros Leontiou}
\affiliation{Department of Mechanical Engineering, Frederick University, 1036 Nicosia, Cyprus}

\author{Stefan Meinel}
\affiliation{Department of Physics, University of Arizona, Tucson, AZ 85721, USA}

\author{Martin Pflaumer}
\affiliation{Goethe-Universit\"at Frankfurt am Main, Institut f\"ur Theoretische Physik, Max-von-Laue-Stra{\ss}e 1, D-60438 Frankfurt am Main, Germany}

\author{Marc Wagner}
\affiliation{Goethe-Universit\"at Frankfurt am Main, Institut f\"ur Theoretische Physik, Max-von-Laue-Stra{\ss}e 1, D-60438 Frankfurt am Main, Germany}
\affiliation{Helmholtz Research Academy Hesse for FAIR, Campus Riedberg, Max-von-Laue-Stra{\ss}e 12, D-60438 Frankfurt am Main, Germany}

\date{April 9, 2024}

\begin{abstract}
We present the first determination of the energy dependence of the $B$-$\bar{D}$ and $B^*$-$\bar{D}$ isospin-0, $S$-wave scattering amplitudes both below and above the thresholds using lattice QCD, which allows us to investigate rigorously whether mixed bottom-charm $\bar{b}\bar{c}ud$ tetraquarks exist as bound states or resonances. The scattering phase shifts are obtained using L\"uscher's method from the energy spectra in two different volumes. To ensure that no relevant energy level is missed, we use large, symmetric $7 \times 7$  and $8 \times 8$ correlation matrices that include, at both source and sink, $B^{(*)}$-$\bar{D}$ scattering operators with the lowest three or four possible back-to-back momenta in addition to local $\bar{b}\bar{c}ud$ operators. We fit the energy dependence of the extracted scattering phase shifts using effective-range expansions. We observe sharp peaks in the $B^{(*)}$-$\bar{D}$ scattering rates close to the thresholds, which are associated with shallow bound states, either genuine or virtual, a few MeV or less below the $B^{(*)}$-$\bar{D}$ thresholds. In addition, we find hints for resonances with masses of order $100$ MeV above the thresholds and decay widths of order $200$ MeV.
\end{abstract}

\maketitle

The majority of experimentally observed mesons can be understood in the quark model as quark-antiquark pairs. However, mesons, which are hadrons with integer spin, can in principle also be composed of two quarks and two antiquarks. The existence of these so-called tetraquarks had already been proposed in the early history of the quark model and QCD \cite{Gell-Mann:1964ewy,Zweig:1964jf-xxx,Jaffe:1976ig}, but clear experimental confirmation was obtained only around a decade ago, for example in form of the observation of the charged $Z_c$ and $Z_b$ states as reviewed in Refs.~\cite{Lebed:2016hpi,Chen:2022asf}. While the masses and decays of the latter strongly indicate the presence of a $\bar c c$ pair or a $\bar b b$ pair, their non-vanishing electric charge implies additionally a light quark-antiquark pair. Recently, there was another experimental breakthrough in the field, namely the detection of the $T_{cc}$ tetraquark with quark flavors $\bar c \bar c u d$ by LHCb \cite{LHCb:2021vvq,LHCb:2021auc}. In contrast to previously observed tetraquarks and tetraquark candidates, its mass is slightly below the lowest meson-meson threshold, making it by far the longest-lived experimentally confirmed tetraquark.
Following the observation of this doubly-charm tetraquark, possible next targets for experimental searches could be mixed bottom-charm tetraquarks with flavor content $\bar{b}\bar{c}ud$. Their production cross section at the LHC is estimated to be about 40 times larger compared to the doubly-bottom $\bar{b}\bar{b}ud$ tetraquark \cite{Ali:2018xfq}. The experimental signatures of a tetraquark are completely different depending on whether its mass is above or below the lowest strong-decay threshold. Thus, reliable theoretical predictions concerning $\bar{b}\bar{c}ud$ tetraquarks are very important and also urgent.

On the theoretical side, for the lightest $\bar{b}\bar{b}ud$ tetraquark with $I(J^P) = 0(1^+)$ (which is the bottom-quark partner of the previously mentioned $T_{cc}$), there is a consensus from recent lattice-QCD calculations that it is deeply bound \cite{Francis:2016hui,Junnarkar:2018twb,Leskovec:2019ioa,Mohanta:2020eed,Meinel:2022lzo,Hudspith:2023loy,Aoki:2023nzp} and will decay through the weak interaction only (see Refs.~\cite{Xing:2018bqt,Ali:2018xfq,Hernandez:2019eox} for discussions of possible decay modes). For the case of $\bar{b}\bar{c}ud$, there is no such consensus. After finding initial hints for a possible QCD-stable $\bar{b}\bar{c}ud$ bound state with $I(J^P) = 0(1^+)$ from lattice QCD \cite{Francis:2018jyb}, the same authors refined their calculation with larger lattice sizes and other improvements, and the hints disappeared \cite{Hudspith:2020tdf}. In Ref.~\cite{Meinel:2022lzo}, some of us also performed lattice-QCD calculations of the $\bar{b}\bar{c}ud$ energy spectra for both $I(J^P) = 0(1^+)$ and $I(J^P) = 0(0^+)$, and we likewise did not find any evidence for QCD-stable bound states (although we could not \emph{rule out} a shallow bound state). In contrast, another independent group very recently reported an $I(J^P) = 0(1^+)$, $\bar{b}\bar{c}ud$ bound state $43\left(^{+7}_{-6}\right)\left(^{+24}_{-14}\right)$ MeV below the $B^*$-$\bar{D}$ threshold based on their lattice-QCD study \cite{Padmanath:2023rdu}, in which the $B^*$-$\bar{D}$ scattering length was determined using the L\"uscher method \cite{Luscher:1986pf,Luscher:1990ck,Luscher:1990ux,Luscher:1991cf} applied to the ground state. Non-lattice approaches also do not show a consistent picture. While Refs.~\cite{Lee:2009rt,Chen:2013aba,Karliner:2017qjm,Sakai:2017avl,Deng:2018kly,Agaev:2018khe,Carames:2018tpe,Yang:2019itm,Tan:2020ldi,Guo:2021yws,Richard:2022fdc,Liu:2023vrk} predict a QCD-stable $\bar{b}\bar{c}ud$ tetraquark, Refs.~\cite{Ebert:2007rn,Eichten:2017ffp,Park:2018wjk,Braaten:2020nwp,Lu:2020rog,Song:2023izj} reached the opposite conclusion.

In the following, we present a new lattice-QCD study of the $\bar{b}\bar{c}ud$ systems with both $I(J^P) = 0(1^+)$ and $I(J^P) = 0(0^+)$. This study uses a different lattice setup and substantially more advanced methods compared to previous work, allowing us to apply the L\"uscher method to multiple excited states in addition to the ground state and hence to reliably determine the detailed energy dependence of the $B$-$\bar{D}$ and $B^*$-$\bar{D}$ isospin-0, $S$-wave scattering amplitudes.

In lattice QCD, the low-lying finite-volume energy levels with a given set of quantum numbers (the total spatial momentum, the quark flavor content, and the irreducible representation of the full octahedral group) are extracted from numerical results for imaginary-time two-point correlation functions $C_{ij}(t) = \langle \op_i(t)\op_j^\dag(0)\rangle$. The operators $\op_i$ are constructed out of quark and gluon fields such that they excite states with the desired quantum numbers, which resemble the low-lying energy eigenstates of interest. For an infinite (in practice, large) time extent of the lattice, the two-point function is equal to  $C_{ij}(t) = \sum_n \langle \Omega| \op_i(0) |n\rangle\langle n|\op_j^\dag(0)|\Omega\rangle\:e^{-E_n t}$, where $|\Omega\rangle$ is the vacuum state and the sum is over all eigenstates $|n\rangle$ of the finite-volume QCD Hamiltonian for which the product of overlap matrix elements is nonzero. By analyzing the time dependence of the numerical results for $C_{ij}(t)$, the energies $E_n$ can be extracted. Because lattice QCD uses a Monte-Carlo sampling of the Euclidean path integral, the numerical results have statistical uncertainties. Moreover, these uncertainties typically grow exponentially with $t$.

For multi-quark systems, experience has shown that the simplest possible operator choices in which the quark fields are combined at the same spacetime point (``local'' operators) are often insufficient to reliably extract even just the ground state \cite{Detmold:2019ghl}. The reason is that all or most of the energy levels resemble multi-hadron states with specific relative momenta, and the spectrum of such states in the case of heavy-quark systems is particularly dense. Among the previous lattice studies of $\bar{b}\bar{c}ud$ systems, Refs.~\cite{Francis:2018jyb,Hudspith:2020tdf,Padmanath:2023rdu} used only local four-quark operators with various types of smearing (local, wall, box) applied to each quark. Reference \cite{Meinel:2022lzo} improved upon this by including also two-meson ($B$-$\bar{D}$ and $B^*$-$\bar{D}$) ``scattering'' operators, that is, operators with each meson individually projected to a specific momentum (equal to zero only, in this case). These operators were included at the sink only, to avoid having to generate expensive all-to-all light-quark propagators. The work presented in the following no longer makes this restriction and is the first lattice-QCD calculation of $\bar{b}\bar{c}ud$ correlation matrices with $B^{(*)}$-$\bar{D}$ scattering operators at both source and sink, and also the first to include $B^{(*)}$-$\bar{D}$ scattering operators with nonzero back-to-back momenta.

Specifically, to study the $\bar{b}\bar{c}ud$ system with $I(J^P) = 0(0^+)$, we use seven operators $\op_{1...7}^{A_1^+}$, of which $\op_{1}^{A_1^+}$ through $\op_{3}^{A_1^+}$ are operators with all four quarks at the same spacetime point (but with Gaussian smearing of the quark fields) and jointly projected to zero total spatial momentum, and $\op_{4}^{A_1^+}$ through $\op_{7}^{A_1^+}$ are $B$-$\bar{D}$ scattering operators with zero total spatial momentum in which the $B$ and $\bar{D}$ operators have back-to-back momenta of magnitudes $0$, $2\pi/L$, $\sqrt{2}\cdot 2\pi/L$, and $\sqrt{3}\cdot 2\pi/L$ ($L$ is the spatial lattice size). Similarly, for the $\bar{b}\bar{c}ud$ system with $I(J^P) = 0(1^+)$, we use eight operators $\op_{1...8}^{T_1^+}$, of which $\op_{1}^{T_1^+}$ through $\op_{4}^{T_1^+}$ are local four-quark operators and $\op_{5}^{T_1^+}$ through $\op_{8}^{T_1^+}$ are $B^*$-$\bar{D}$ scattering operators in which the $B^*$ and $\bar{D}$ have back-to-back momenta of magnitudes $0$, $2\pi/L$ (for both $\op_{6}^{T_1^+}$ and $\op_{7}^{T_1^+}$), and $\sqrt{2}\cdot 2\pi/L$. Two different operators are used for the case with one unit of back-to-back momentum to account for the mixing of $S$ and $D$ partial waves \cite{Woss:2018irj}. The labels $A_1^+$ and $T_1^+$ refer to the octahedral-group irreps of positive parity that contain the angular momenta $J=0,4,...$ and $J=1,3,...$, respectively. The explicit definitions of all operators are given in the supplemental material. We compute the symmetric $7\times7$ and $8\times8$ correlation matrices of these operators, using combinations of (Gaussian smeared) point-to-all and stochastic timeslice-to-all propagators \cite{Abdel-Rehim:2017dok}.

Our calculations were performed for two different lattice sizes, $24^3\times 64$ and $32^3\times 64$, with a lattice spacing of approximately 0.12 fm and pion mass of approximately 220 MeV in both cases. The charm and bottom valence quarks were implemented using the Fermilab method \cite{El-Khadra:1996wdx} and lattice NRQCD \cite{Lepage:1992tx}, respectively, with approximately physical kinetic masses. Further details are provided in the supplemental material.

We computed the $\bar{b}\bar{c}ud$ correlation matrices for approximately 1000 gauge configurations on each lattice, with 30 source locations per configuration for the elements computed using (Gaussian smeared) point-source propagators, and 3 random $\mathbb{Z}_2\times\mathbb{Z}_2$ sources on 4 timeslices per configuration for the elements computed with (Gaussian smeared) stochastic propagators; we also use color and spin dilution and the one-end trick \cite{Abdel-Rehim:2017dok}. To extract the $\bar{b}\bar{c}ud$ finite-volume energy levels from these correlation matrices, we follow the well-established approach of solving the generalized eigenvalue problem (GEVP) \cite{Luscher:1990ck,Blossier:2009kd}
\begin{align}
 \sum_j C_{ij}(t) v_{j,n}(t,t_0) = \lambda_n(t,t_0) \sum_j C_{ij}(t_0) v_{j,n}(t,t_0),
\end{align}
where we set $t_0/a=3$ and verified that the results do not significantly depend on this choice. We then perform single-exponential fits of the form $\lambda_n(t,t_0)=A_n e^{-E_n t}$ to obtain the energy levels $E_n$; see the supplemental material for further details.

\begin{figure}
 \includegraphics[width=\linewidth]{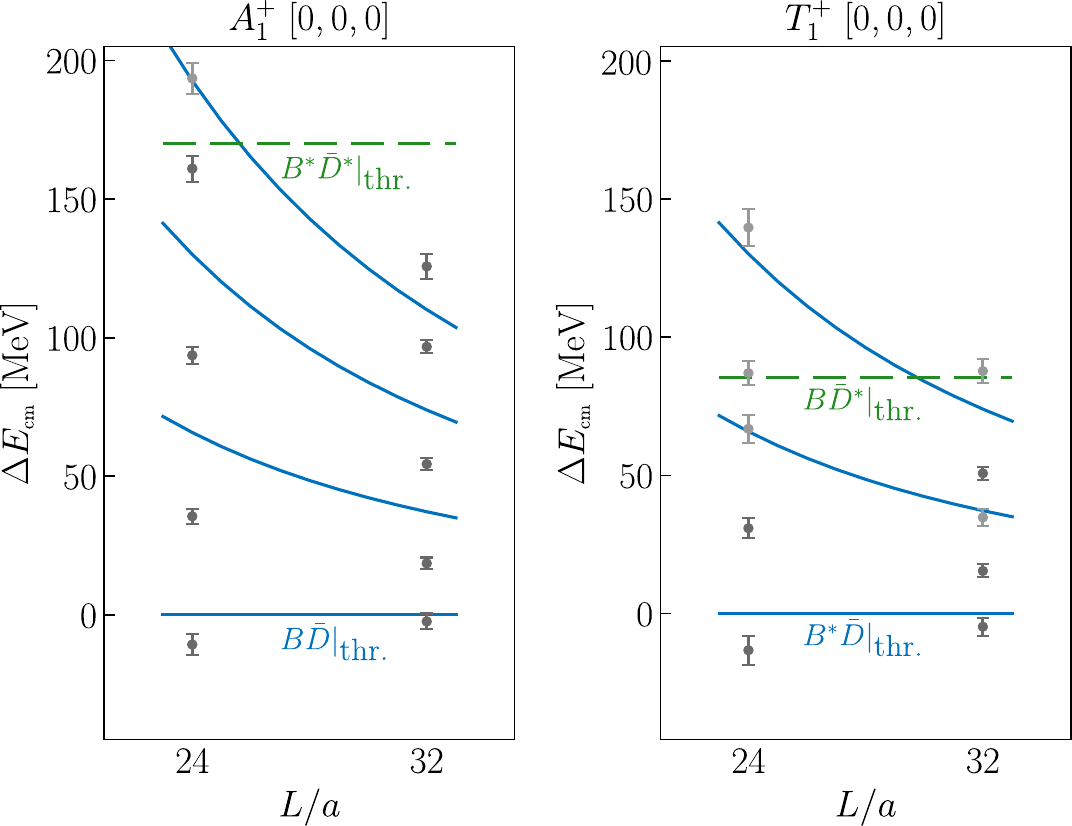}
 \caption{\label{fig:EvsL}{\it Left:} The finite-volume energies of the $\bar{b}\bar{c}ud$ system with $I(J^P) = 0(0^+)$ as a function of the spatial lattice size.  The data points with error bars show the actual finite-volume energy levels; points plotted with a lighter shade of gray are excluded from the L\"uscher analysis. The solid blue curves correspond to what would be the noninteracting $B$-$\bar{D}$ energy levels, the lowest of which coincides with the strong-decay threshold. The dashed green line shows the $B^*$-$\bar{D}^*$ threshold. {\it Right:} The corresponding plot for $I(J^P) = 0(1^+)$. Here, the solid blue curves correspond to what would be the noninteracting $B^*$-$\bar{D}$ energy levels, and the dashed green line corresponds to the $B$-$\bar{D}^*$ threshold. }
\end{figure}

Our results for the lowest five energy levels of each $\bar{b}\bar{c}ud$ system are shown as a function of the spatial lattice size $N_s=L/a$ in Fig.~\ref{fig:EvsL}. Also shown are the lowest four noninteracting $B^{(*)}$-$\bar{D}$ energy levels, calculated as $E=E_{B^{(*)}}(\mathbf{p}^2)+E_{\bar{D}}(\mathbf{p}^2)$ with momenta $\mathbf{p}$ satisfying the periodic boundary conditions [each component an integer multiple of $2\pi/L$], and with the single-meson energies calculated on the lattice and described by the dispersion relations
\begin{align}
\nonumber E_{B^{(*)}}(\mathbf{p}^2)=&E_{B^{(*)}}(0)+\sqrt{m_{B^{(*)},{\rm kin}}^2+\mathbf{p}^2}-m_{B^{(*)},{\rm kin}}, \\
 E_D(\mathbf{p}^2)=&E_{\bar{D}}(0)+\mathbf{p}^2/(2m_{\bar{D},{\rm kin}})-\mathbf{p}^4/(8m_{\bar{D},4}^3). \label{eq:disprels}
\end{align}
The values of $E_{B^{(*)}}(0)$,\: $m_{B^{(*)},{\rm kin}}$,\: $E_{\bar{D}}(0)$,\: $m_{\bar{D},{\rm kin}}$, and $m_{\bar{D},4}$ are provided in the supplemental material. In Fig.~\ref{fig:EvsL} we see that the actual $\bar{b}\bar{c}ud$ energy levels are shifted significantly relative to the noninteracting levels due to the meson-meson interactions in the finite volume, except for the third level in the case of $J=1$ (we discuss the reason for this behavior farther below). 

To rigorously investigate whether bound states or resonances exist, we map the observed finite-volume energy levels $E_n$ to infinite-volume $S$-wave $B^{(*)}$-$\bar{D}$ scattering phase shifts $\delta_0(k_n)$ using the L\"uscher quantization condition
\begin{align}
\label{eq:Luscher} \cot{\delta_0(k_n)} = \frac{2 Z_{00}(1;(k_n L/2 \pi)^2)}{\pi^{1/2} k_n L} ,
\end{align}
where $Z_{00}$ is the generalized zeta function \cite{Luscher:1990ux} and $k_n$ is the scattering momentum associated with energy level $E_n$, calculated from $E_n=E_{B^{(*)}}(k_n^2)+E_{\bar{D}}(k_n^2)$ with the dispersion relations (\ref{eq:disprels}). To ensure that the single-channel, single-partial-wave approximation is applicable, we only extract the phase shifts for the energy levels below the $B^*$-$\bar{D}^*$ ($J=0$) and $B$-$\bar{D}^*$ ($J=1$) thresholds. Furthermore, for $J=1$, we observe that the third finite-volume energy level is consistent with the noninteracting $|\mathbf{p}|=2\pi/L$ energy level that has multiplicity 2 once including both $S$-wave and $D$-wave structures, as we did in our operator basis. Because finite-volume interactions for higher partial waves are suppressed, we conclude that this energy level is dominantly $D$-wave, and we therefore exclude it from the L\"uscher analysis. This is further corroborated by the eigenvectors from the GEVP, which show that this state has a non-negligible overlap only with the operator $\op^{T_1^+}_{7}$ that was subduced from a $D$-wave structure. Note that there are similar degeneracies for higher non-interacting levels above the energy region used in our analysis \cite{Woss:2018irj}.
 
\begin{table}
	\centering
	\begin{tabular}{lcccccc} \hline \hline
	& &	$\Delta m_{\rm GBS}$ [MeV] & & $\Delta m_{\textrm{R}}$ [MeV] & & $\Gamma_{\textrm{R}}$ [MeV]   \\ \hline 
	\vspace{-0.3cm} \\
	$J\!=\!0$ & &	$-0.5_{-1.5}^{+0.4}$ & & $ 138(13) $ & & $ 229(35)$  \\ 
	\vspace{-0.3cm} \\
	$J\!=\!1$ & &	$-2.4_{-0.7}^{+2.0}$ & & $ \phantom{0}67(24) $ & & $ 132(32)$  \\
	\vspace{-0.3cm} \\ \hline \hline
	\end{tabular}
	\caption{\label{tab:finalresults}Our results for the $\bar{b}\bar{c}ud$ genuine-bound-state (GBS) and resonance (R) pole locations, where $\Delta m_{\rm GBS}=m_{\rm GBS}-m_{B^{(*)}}-m_{\bar{D}}$, $\Delta m_{\textrm{R}}=\mathrm{Re}(\sqrt{s_{\textrm{R}}})-m_{B^{(*)}}-m_{\bar{D}}$, and $\Gamma_{\textrm{R}}=-2\:\mathrm{Im}(\sqrt{s_{\textrm{R}}})$. Only the statistical uncertainties are given.}
\end{table}

\begin{figure*}

\parbox{0.55\linewidth}{
 \includegraphics[width=\linewidth]{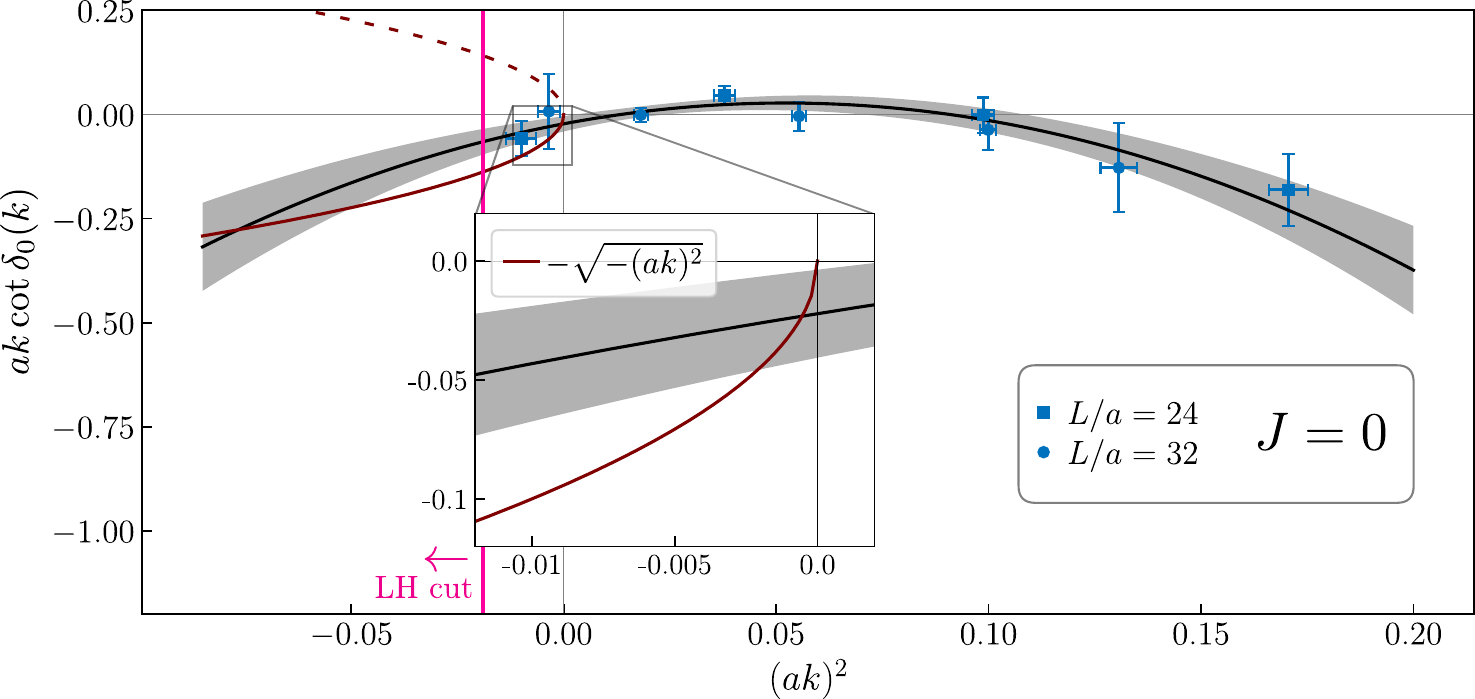}
 
 \vspace{1ex}

 \includegraphics[width=\linewidth]{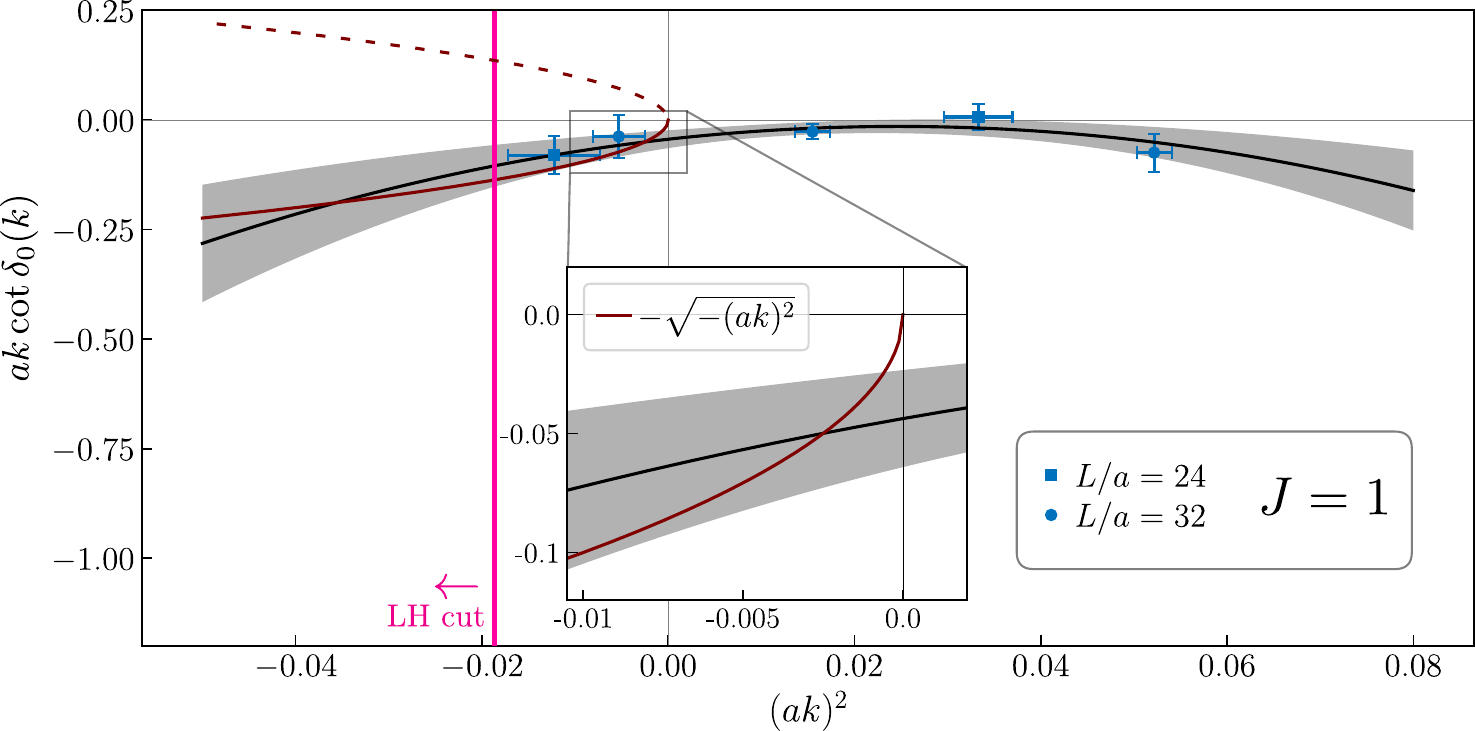}
}
\hfill
\parbox{0.41\linewidth}{
 \includegraphics[width=\linewidth]{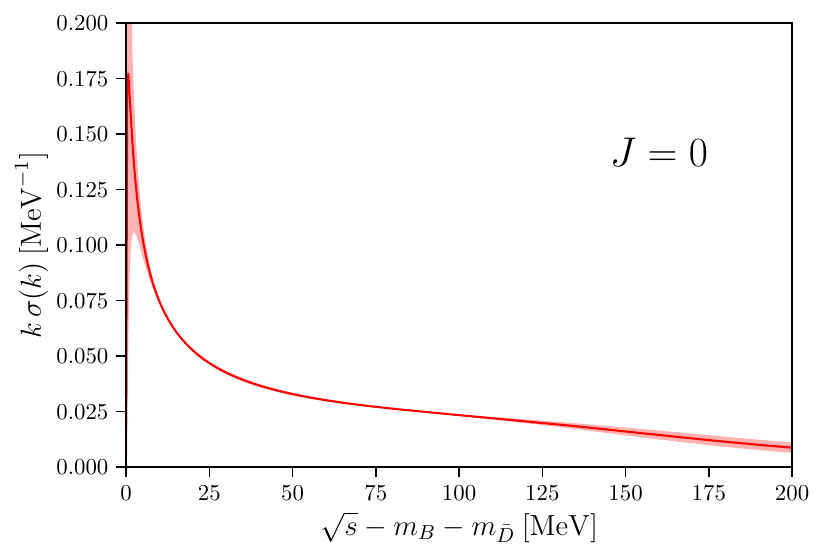}

 \includegraphics[width=\linewidth]{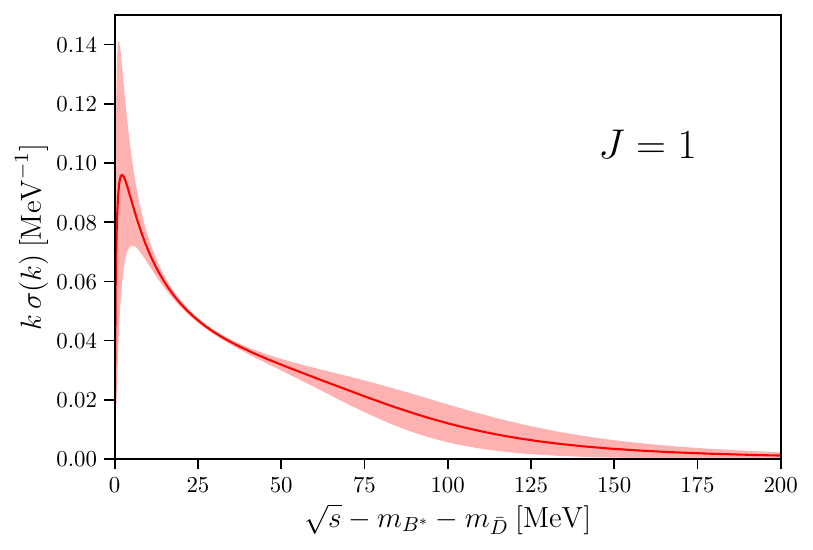}
}

\vspace{-1ex}
 
 \caption{\label{fig:phasehiftsandpoles}{\it Left:} Our results for the functions $ak \cot\delta_0(k)$ for $S$-wave $B$-$\bar{D}$ scattering (top) and $S$-wave $B^*$-$\bar{D}$ scattering (bottom), where $k$ is the scattering momentum, $\delta_0(k)$ is the scattering phase shift, and $a=0.11887(80)$ fm is the lattice spacing. The data points were obtained directly from the lattice energy levels, and the curves correspond to ERE fits through order $k^4$. Also shown are the functions $-\sqrt{-(ak)^2}$ (solid red parabolas) whose intersections with $ak \cot\delta_0(k)$ just below threshold correspond to the shallow $\bar{b}\bar{c}ud$ bound states we predict, and  $+\sqrt{-(ak)^2}$ (dashed red parabolas) whose intersections with $ak \cot\delta_0(k)$ would correspond to virtual $\bar{b}\bar{c}ud$ bound states. The vertical magenta lines show the positions of two-pion-exchange left-hand branch points. {\it Right:} Our results for the product of scattering momentum and $B^{(*)}$-$\bar{D}$ scattering cross section, which is proportional to the scattering rate, as a function of center-of-momentum energy.}
\end{figure*}

Our results for the scattering phase shifts, along with effective-range expansion (ERE) fits of the form 
	\begin{equation}
		k \cot \delta_0(k) = \frac{1}{a_0} + \frac{1}{2}r_0 k^2 + b_0 k^4,
		\label{eq:Scatt_ERE_Quad}
	\end{equation}
are shown in Fig.~\ref{fig:phasehiftsandpoles} (Left). The numerical values of the fitted ERE parameters are given in the supplemental material. The scattering phase shift is related to the $S$-wave scattering amplitude and cross section by
\begin{align}
T_0(k) = \frac{1}{\cot\delta_0(k) - i}, \hspace{4ex} \sigma(k)=\frac{4\pi}{k^2}|T_0(k)|^2. \label{eq:scatteringamp}
\end{align}
Poles of $T_0(k)$ at purely imaginary $k$ correspond to genuine or virtual bound states for $\mathrm{Im}(k)>0$ or $\mathrm{Im}(k)<0$, respectively, while poles with $\mathrm{Re}(k)\neq0$ and $\mathrm{Im}(k)<0$ correspond to resonances. Using our ERE fits, we find genuine bound-state poles as well as resonance poles for both $J=0$ and $J=1$ at the values of $\sqrt{s}-\sqrt{s_{\rm th}} = \sqrt{m_{B^{(\ast)}}^2 + k^2} + \sqrt{m_{\bar{D}}^2 + k^2}-m_{B^{(*)}}-m_{\bar{D}}$ given in Table~\ref{tab:finalresults} ($s$ denotes the Mandelstam variable equal to the square of the center-of-momentum energy). We used our lattice results for the kinetic $B^{(*)}$ and $\bar{D}$ masses to evaluate this expression; to obtain predictions for absolute tetraquark bound state or resonance masses, one simply needs to add the experimental value of the threshold energy, $m_{B^{(*)}}^\text{exp} + m_{\bar{D}}^\text{exp}$.

The resonances have masses of order $100$ MeV above the $B^{(*)}$-$\bar{D}$ thresholds and decay widths of order $200$ MeV. We caution that the resonance poles lie outside the radius of convergence of the ERE, which is limited by the presence of a left-hand cut associated with two-pion $t$-channel exchange in the scattering process (single-pion exchange would require a $D^*$ in the initial or final state, and is therefore not relevant here). The center-of-momentum energy at which the left-hand cut starts is obtained from the kinematic relations for the Mandelstam variables by expressing $s$ in terms of $t$ and the scattering angle $\theta^*$, and then setting $t=(2m_\pi)^2$ and $\theta^*=\pi$ \cite{Raposo:2023nex}; this gives $\sqrt{s_{\rm cut}}-\sqrt{s_{\rm th}}\approx-18\:{\rm MeV}$ for both $J = 0$ and $J = 1$, corresponding to $(ak)^2_{\rm cut}\approx -0.019$, as indicated with the magenta lines in Fig.~\ref{fig:phasehiftsandpoles}. While our ERE fit is seen to describe the data very well for real $(ak)^2$ in the full momentum range, the prediction of resonance poles away from the real axis may be less reliable. We note, however, that instead of expanding $k \cot \delta_0(k)$ around $k^2 = 0$ one could as well expand around the midpoint of the left hand cut and the second threshold. The convergence radius in the complex energy plane would then be significantly larger, while the resulting parametrizations of the lattice data points for $k \cot \delta_0(k)$ would be identical to those obtained from the ERE. The reason is that in both cases second-order polynomials in $k^2$ are fitted to the same data points. This suggests that our results can be trusted in a significantly larger region of the complex energy plane, namely disks of radius $94 \, \text{MeV}$ around $E = 76 \, \text{MeV}$ for $J = 0$ and of radius $52 \, \text{MeV}$ around $E = 34 \, \text{MeV}$ for $J = 1$. The predicted resonances are still located outside, but quite close to the boundaries of these convergence regions. To further test their stability, we also performed ERE fits through order $k^6$. The coefficients of $k^6$ are found to be consistent with zero within the statistical uncertainties, and the other parameters remain consistent with those from the order-$k^4$ fit. For $J=0$, the resonance pole obtained from the order-$k^6$ fit is at a similar location. For $J=1$, where we have fewer data points, the uncertainties from the $k^6$ fit are too large to determine the pole locations.

The bound-state poles are extremely close to threshold and therefore well within the region of validity of the ERE. However, as can be seen from the $\pm\sqrt{-(ak)^2}$ parabolas in Fig.~\ref{fig:phasehiftsandpoles}, statistical fluctuations could turn the genuine bound states into virtual bound states, which are not asymptotic states in QCD but would still strongly affect the $B^{(*)}$-$\bar{D}$ scattering rates near threshold \cite{Padmanath:2022cvl}, or could, at first glance, lead to the complete disappearance of the poles.
To quantify this uncertainty, we generated $10 \, 000$ multivariate Gaussian random samples for the ERE fit parameters $a_0$, $r_0$, and $b_0$ according to their mean values and covariance matrix. For each sample, we first checked whether it is still consistent with our finite-volume energy spectra. To this end, we calculate the two finite-volume energy spectra predicted by the sample using the L\"uscher quantization condition. For some of the samples, the intersection of $k \cot \delta_0(k)$ and $2 Z_{00}(1;(k_n L/2 \pi)^2)/(\pi^{1/2} L)$ below threshold disappears for at least one of the volumes, leaving only the intersections that match the first and higher excited-state energy levels calculated on the lattice. These samples are therefore inconsistent with the observed spectra and we removed them. For each of the remaining samples, we then checked whether there is a genuine bound state (GBS), a virtual bound state (VBS), or no bound state (NBS), with the following outcome:
\begin{itemize}
\item $J = 0$: $88.5\%$ GBS, $11.5\%$ VBS, $0.0\%$ NBS,
\item $J = 1$: $97.7\%$ GBS, $2.3\%$ VBS, $0.0\%$ NBS.
\end{itemize}
The lower and upper limits for $\Delta m_{\rm GBS}$ given in Table \ref{tab:finalresults} correspond to the 16th and 84th percentiles of those random samples for which genuine bound states exist, while the central values correspond to the best-fit points. To further test our prediction of shallow bound states, we performed additional ERE fits of order $k^0$ and order $k^2$ using only the three data points closest to threshold, which are within the strict radius of convergence of the ERE. These fits, which are shown in the supplemental material, give consistent results.

Returning to the discussion of our main fits as shown in Fig.~\ref{fig:phasehiftsandpoles}, we note that, in addition to the shallow-bound-state and the broad-resonance poles, there are poles with purely imaginary $k$ below the left-hand branch point. These poles must be discarded because the direction of the crossing corresponds to a pole residue with an unphysical sign \cite{Iritani:2017rlk} and our parametrization of $k \cot \delta_0(k)$ is invalid in that region (bound states this far below threshold are also ruled out by the absence of corresponding finite-volume energy levels).

The scattering rate (probability per time) is equal to the product of flux and cross section, and hence proportional to $k\,\sigma(k)$ for nonrelativistic $k$. These products are shown in Fig.~\ref{fig:phasehiftsandpoles} (Right) as a function of the center-of-momentum energy. We emphasize that the scattering rates only depend on our fit functions for real-valued $k^2$ that interpolate our data very well, so these predictions are also expected to be very reliable. We observe sharp enhancements in the scattering rates close to the thresholds, related to the shallow bound states or virtual bound states. At higher energies, the scattering rates continue to be enhanced, likely by the broad resonances. The scattering rates are very close to the largest possible value allowed by unitarity, given by $|T_0|^2=1$, up to several tens of MeV above threshold.

In summary, the substantial improvements made here in determining the $\bar{b}\bar{c}ud$ finite-volume energy levels allowed us to determine the detailed energy dependence of the $B$-$\bar{D}$ and $B^*$-$\bar{D}$ $S$-wave scattering amplitudes for the first time using lattice QCD, revealing very interesting strong-interaction phenomena. We found poles for both $J=0$ and $J=1$ corresponding to shallow bound states, as well as hints for poles corresponding to broad resonances. While further lattice-QCD computations at additional lattice spacings and pion masses will be needed to pin down the exact location and nature of each pole at the physical point, we expect our prediction of shallow bound states, either genuine or virtual, to be quite robust. The possible resonances above threshold are very broad and are therefore presumably difficult to observe at the LHC and future experiments. On the other hand, if the $J=0$ pole just below the $B$-$\bar{D}$ threshold is confirmed as a genuine bound state, this isoscalar, scalar $\bar{b}\bar{c}ud$ tetraquark will decay through the weak interaction only and could become the first tetraquark to be observed at the LHC with this feature. If the $J=1$ pole just below the $B^*$-$\bar{D}$ threshold is confirmed as a genuine bound state, it will decay electromagnetically into $B \bar{D} \gamma$ (and also into the $J=0$ tetraquark plus a photon, if that tetraquark is confirmed as a genuine bound state).

\begin{acknowledgments}
\textit{Acknowledgments:} We thank Simone Bacchio, Luka Leskovec, Mikhail Mikhasenko, Akaki Rusetsky, Christopher Thomas, Bira van Kolck, and David Wilson for helpful discussions. We thank the MILC collaboration, and in particular Doug Toussaint, for providing the gauge-link ensembles. C.A.\ acknowledges partial support by the project 3D-nucleon, ID number EXCELLENCE/0421/0043, co-financed by the European Regional Development Fund and the Republic of Cyprus through the Research and Innovation Foundation. J.F.\ received financial support by the German Research Foundation (DFG) research unit FOR5269 ``Future methods for studying confined gluons in QCD,'' by the PRACE Sixth Implementation Phase (PRACE-6IP) program (grant agreement No.~823767) and by the EuroHPC-JU project EuroCC (grant agreement No.~951740) of the European Commission. S.M.\ is supported by the U.S. Department of Energy, Office of Science, Office of High Energy Physics under Award Number D{E-S}{C0}009913. M.W.\ and M.P.\ acknowledge support by the Deutsche Forschungsgemeinschaft (DFG, German Research Foundation) -- project number 457742095. M.W.\ acknowledges support by the Heisenberg Programme of the Deutsche Forschungsgemeinschaft (DFG, German Research Foundation) -- project number 399217702. We gratefully acknowledge the Cyprus Institute for providing computational resources on Cyclone under the project IDs p054 and p147. Calculations were also conducted on the GOETHE-HLR and on the FUCHS-CSC high-performance computers of the Frankfurt University. We would like to thank HPC-Hessen, funded by the State Ministry of Higher Education, Research and the Arts, for programming advice. This project also used resources at NERSC, a DOE Office of Science User Facility at LBNL. The computations utilized the QUDA library \cite{Clark:2009wm,Babich:2011np,Clark:2016rdz}.
\end{acknowledgments}

\providecommand{\href}[2]{#2}\begingroup\raggedright\endgroup

\appendix
\onecolumngrid
\newpage

\section*{Supplemental material}
\renewcommand \thesubsection{\Roman{subsection}}
\FloatBarrier

\subsection{Lattice actions and parameters}

Our calculation uses the mixed-action setup that was tested and used successfully by the PNDME collaboration for nucleon-structure computations \cite{Bhattacharya:2015wna,Gupta:2018qil}. This setup employs gauge configurations generated with 2+1+1 flavors of highly improved staggered (HISQ) sea quarks by the MILC collaboration \cite{MILC:2012znn}, but uses the clover-improved Wilson action with HYP-smeared gauge links for the valence light quarks. Here we include two ensembles that differ only in the spatial lattice extent; their main properties are given in Table \ref{tab:configurations}. We set the bare valence light-quark mass to $am_l=-0.075$ and the clover coefficient to $c_{\rm SW}=1.05091$ \cite{Bhattacharya:2015wna,Gupta:2018qil}. We implement the valence charm quarks with the same form of clover-Wilson action and same value of $c_{\rm SW}$, but with mass parameter tuned according to the Fermilab method to eliminate the main heavy-quark discretization errors \cite{El-Khadra:1996wdx}. That is, the bare mass is tuned such that the spin-averaged kinetic $\bar{D}$-meson mass $m_{\bar{D},\textrm{kin}}^{\textrm{spinav}} = (m_{\bar{D},\textrm{kin}}+ 3m_{\bar{D}^*,\textrm{kin}} )/4 $ matches its experimental value \cite{ParticleDataGroup:2022pth}; this condition is satisfied for our final choice $am_c=0.6835775$ at the 3\% level. For the valence bottom quarks, we use order-$v^4$ lattice NRQCD \cite{Lepage:1992tx} with tadpole improvement and order-$\alpha_s$ corrections to the matching coefficients for the kinetic terms; all parameters are given in Ref.~\cite{HPQCD:2011qwj}. The resulting spin-averaged kinetic $B$-meson mass is within 4\% of the experimental value \cite{ParticleDataGroup:2022pth} (see Sec.~\ref{sec:disprels} below for details).

\begin{table}[h]
	\centering
	\begin{tabular}{lcccc}\hline \hline
		Ensemble & $ N_s^3\times N_t $ & $ a $ [fm] & $ m_{\pi}^{(\textrm{sea})} $ [MeV] & $ m_{\pi}^{(\text{val})} $ [MeV]  \\
		\hline
		a12m220S & $ 24^3 \times 64 $ & 0.1202(12) & $ 218.1(4) $ & $ 225.0(2.3) $ \\
		a12m220  & $ 32^3 \times 64 $ & 0.1184(10) & $ 216.9(2) $ & $ 227.9(1.9) $ \\
 \hline \hline& 
	\end{tabular}
	\caption{\label{tab:configurations}The main properties of the two gauge-link ensembles used in this work. Here, $N_s$ and $N_t$ are the numbers of lattice sites in spatial and temporal directions, $a$ is the lattice spacing from the $r_1$ scale \cite{MILC:2012znn}, $m_{\pi}^{(\rm sea)}$ is the mass of the lightest pion formed by the HISQ sea quarks (scale set using $f_{p4s}$) \cite{MILC:2012znn}, and $ m_{\pi}^{(\rm val)}$ is the mass of the pion constructed with the clover-Wilson valence quarks \cite{Bhattacharya:2015wna,Gupta:2018qil}. Because the bare action parameters of both ensembles are identical, we use the weighted average lattice spacing $a=0.11887(80)$ fm for both ensembles.}
\end{table}

\subsection{Operators}

For the $\bar{b}\bar{c}ud$ system with $I(J^P) = 0(0^+)$, we use the seven operators
\begin{alignat}{2}
		\label{eq:op_MF_BD_total_zero} &\op_1^{A_1^+} =&& \frac{1}{\sqrt{V_S}} \sum_{\mathbf{x}}	\bar{b}(\mathbf{x})\g5 u(\mathbf{x}) \, \bar{c}(\mathbf{x})\g5 d(\mathbf{x}) - (d \leftrightarrow u) ,	 \\
		\label{eq:op_MF_BastDast_total_zero} &\op_2^{A_1^+} =&& \frac{1}{\sqrt{V_S}}\sum_{\mathbf{x}} \bar{b}(\mathbf{x})\g{j} u(\mathbf{x}) \, \bar{c}(\mathbf{x})\g{j} d(\mathbf{x}) - (d \leftrightarrow u), \\
		&\op_3^{A_1^+} =&&\frac{1}{\sqrt{V_S}} \sum_{\mathbf{x}}\bar{b}^a(\mathbf{x}) \g5 C \bar{c}^{b,T}(\mathbf{x})\, u^{a,T}(\mathbf{x})  C \g5  d^b(\mathbf{x}) - (d \leftrightarrow u), \label{eq:op_MF_Dd_bcud_J0_total_zero} \\
	\label{eq:op_bcud_J0_CoM_1}	&\op^{A_1^+}_{4} =  &&\;	B^+(\mathbf{0}) \, D^-(\mathbf{0}) - (d \leftrightarrow u),  \vphantom{\sum_x}\\
	\label{eq:op_bcud_J0_CoM_2}	&\op^{A_1^+}_{5} =  &&\sum_{\mathbf{q} =\pm \mathbf{e}_{i=x,y,z}} 
	B^+( \mathbf{q}) \, D^-(- \mathbf{q}) - (d \leftrightarrow u), \\
	\label{eq:op_bcud_J0_CoM_3}	&\op^{A_1^+}_{6} = &&\sum_{\mathbf{q} = \pm \mathbf{e}_{i} \pm \mathbf{e}_{j},\, i<j}
	B^+(\mathbf{q}) \, D^-(- \mathbf{q}) - (d \leftrightarrow u) ,\\
	\label{eq:op_bcud_J0_CoM_4}	&\op^{A_1^+}_{7} = &&\sum_{\mathbf{q} = \pm \mathbf{e}_{x} \pm \mathbf{e}_{y} \pm \mathbf{e}_{z}}
	B^+(\mathbf{q}) \, D^-(- \mathbf{q}) - (d \leftrightarrow u),
\end{alignat}
where the repeated index $j$ is summed over the spatial directions, the repeated color indices $a$ and $b$ are summed over the three colors, $V_S=L^3$ is the spatial lattice volume, and
\begin{align}
	B^{+}(\mathbf{q}) =& \frac{1}{\sqrt{V_S}} \sum_{ \mathbf{x}} \,\bar{b}(\mathbf{x}) \g5 u(\mathbf{x}) \, e^{i \frac{2\pi}{L}\mathbf{q}\cdot \mathbf{x}}, \\
	D^{-}(\mathbf{q}) =& \frac{1}{\sqrt{V_S}} \sum_{ \mathbf{x}} \,\bar{c}(\mathbf{x}) \g5 d(\mathbf{x})  \, e^{i \frac{2\pi}{L}\mathbf{q}\cdot \mathbf{x}}. \label{eq:Dop}
\end{align}
The operators $\op_1^{A_1^+}$ and $\op_2^{A_1^+}$ are constructed as products of color-singlet $B$, $\bar{D}$ and $B^*$, $\bar{D}^*$ operators at the same spacetime point that are then jointly projected to zero momentum by summing over the spatial coordinates. The operator $\op_3^{A_1^+}$ is constructed as a color-singlet contraction of two color-nonsinglet diquarks at the same spacetime point that is then jointly projected to zero momentum. The operators $\op_{4}^{A_1^+}$ through $\op_{7}^{A_1^+}$ are $B$-$\bar{D}$ ``scattering'' operators in which the $B$ and $\bar{D}$ operators are individually momentum-projected and have back-to-back momenta of magnitudes $0$, $2\pi/L$, $\sqrt{2}\cdot 2\pi/L$, and $\sqrt{3}\cdot 2\pi/L$ ($L$ is the spatial lattice size). For the scattering operators, the summations over the back-to-back momentum directions ensure that the operators transform in the $A_1^+$ irrep of the octahedral group that contains $J=0$.

All quark fields in the above expressions are smeared using gauge-covariant Gaussian smearing (see e.g.\ Eq.\ (8) in Ref.\ \cite{Leskovec:2019ioa}), with $(\sigma_\text{Gauss}, N_\textrm{Gauss})=(4.47,35),(1.195, 5),(1.0, 10)$ for the light, charm, and bottom quarks, respectively. The gauge links used for the Gaussian smearing are APE smeared (see e.g.\ Eq.\ (23) in Ref.\ \cite{Jansen:2008si}) with parameters $N_\textrm{APE} = 50$ and $\alpha_\textrm{APE} = 0.5$. The smearing parameters are identical at source and sink, leading to symmetric correlation matrices.

For the $\bar{b}\bar{c}ud$ system with $I(J^P) = 0(1^+)$, we use the eight operators
\begin{alignat}{2}
	\label{eq:op_MF_BastD_total_zero}&\op_{1,k}^{T_1^+} =&& \frac{1}{\sqrt{V_S}} \, \sum_{\mathbf{x}} \bar{b}(\mathbf{x})\g{k} u(\mathbf{x}) \, \bar{c}(\mathbf{x})\g5 d(\mathbf{x}) - (d \leftrightarrow u) ,	 \\
	\label{eq:op_MF_BDast_total_zero}&\op_{2,k}^{T_1^+} =&& \frac{1}{\sqrt{V_S}} \, \sum_{\mathbf{x}} \bar{b}(\mathbf{x})\g5 u(\mathbf{x}) \, \bar{c}(\mathbf{x})\g{k} d(\mathbf{x}) - (d \leftrightarrow u) ,	 \\
	\label{eq:op_MF_BastDast_J1_total_zero}&\op_{3,k}^{T_1^+} =&& \frac{1}{\sqrt{V_S}} \, \, \epsilon_{kjl} \sum_{\mathbf{x}} \bar{b}(\mathbf{x})\g{j} u(\mathbf{x}) \, \bar{c}(\mathbf{x})\g{l} d(\mathbf{x})   - (d \leftrightarrow u), \\
	 &\op_{4,k}^{T_1^+} =&& \frac{1}{\sqrt{V_S}} \, \sum_{\mathbf{x}}\bar{b}^a(\mathbf{x}) \g{k} C \bar{c}^{b,T}(\mathbf{x})\, u^{a,T}(\mathbf{x})  C \g5  d^b(\mathbf{x}) - (d \leftrightarrow u), \label{eq:op_MF_Dd_J1_total_zero} \\
	\label{eq:op_bcud_J1_CoM_1}	&\op^{T_1^+}_{5,k} =  &&\;	B_k^{*+}(\mathbf{0}) \, D^-(\mathbf{0}) - (d \leftrightarrow u),  \vphantom{\sum_x}  \\
	\label{eq:op_bcud_J1_CoM_2}	&\op^{T_1^+}_{6,k} =  &&\sum_{\mathbf{q}=\pm \mathbf{e}_{i=x,y,z}} 
	B_k^{*+}( \mathbf{q}) \, D^-(- \mathbf{q}) - (d \leftrightarrow u),  \\
	&\op^{T_1^+}_{7,z} = &&\sum_{\mathbf{q}=\pm \mathbf{e}_{i=x,y}}
	B_{z}^{*+}(\mathbf{q}) \, D^-(- \mathbf{q})
	- 2\sum_{\mathbf{q}'=\pm \mathbf{e}_{z}} B_{z}^{*+}(\mathbf{q}') \, D^-(- \mathbf{q}') - (d \leftrightarrow u), \label{eq:op_bcud_J1_CoM_3}\\
	\label{eq:op_bcud_J1_CoM_4}	&\op^{T_1^+}_{8,k} = &&\sum_{\mathbf{\mathbf{q}} = \pm \mathbf{e}_{i} \pm \mathbf{e}_{j}, i<j}
	B_k^{*+}(\mathbf{q}) \, D^-(- \mathbf{q}) - (d \leftrightarrow u), 
\end{alignat}
where $D^-(\mathbf{q})$ was defined in Eq.~(\ref{eq:Dop}),
\begin{align}
 	B_k^{*+}(\mathbf{q}) = \frac{1}{\sqrt{V_S}} \sum_{\mathbf{x}} \,\bar{b}(\mathbf{x}) \g{k} u(\mathbf{x}) \, e^{i\frac{2\pi}{L}\mathbf{q}\cdot{\mathbf{x}}},
\end{align}
and $k=x,y,z$ denotes the spatial polarization direction (the operator $\op^{T_1^+}_{7}$ is shown for $k=z$ only). These operators transform in the $T_1^+$ irrep of the octahedral group that contains $J=1$. The operators $\op_1^{T_1^+}$, $\op_2^{T_1^+}$, and $\op_3^{T_1^+}$ are constructed as products of color-singlet $B^{(*)}$ and $\bar{D}^{(*)}$ operators at the same spacetime point that are then jointly projected to zero momentum by summing over the spatial coordinates. The operator $\op_4^{T_1^+}$ is constructed as a color-singlet contraction of two color-nonsinglet diquarks at the same spacetime point that is then jointly projected to zero momentum. Here, the two heavy quarks are combined to a flavor-symmetric spin-1 diquark and the two light quarks are combined to a flavor-antisymmetric spin-0 diquark. The operators $\op_{5}^{T_1^+}$ through $\op_{8}^{T_1^+}$ are $B^*$-$\bar{D}$ scattering operators in which the $B^*$ and $\bar{D}$ have back-to-back momenta of magnitudes $0$, $2\pi/L$ (for both $\op_{6}^{T_1^+}$ and $\op_{7}^{T_1^+}$), and $\sqrt{2}\cdot 2\pi/L$. Two different operators are used for the case with one unit of back-to-back momentum to account for the mixing of $S$ and $D$ partial waves \cite{Woss:2018irj}. Again, all quark fields are smeared, with the same parameters as used for $J=0$.

\FloatBarrier
\subsection{\label{sec:disprels}\texorpdfstring{$\bm{\bar{D}^{(*)}}$}{Dbar(*)} and \texorpdfstring{$\bm{B^{(*)}}$}{B(*)} dispersion relations}

\begin{figure}
 \includegraphics[width=0.5\linewidth]{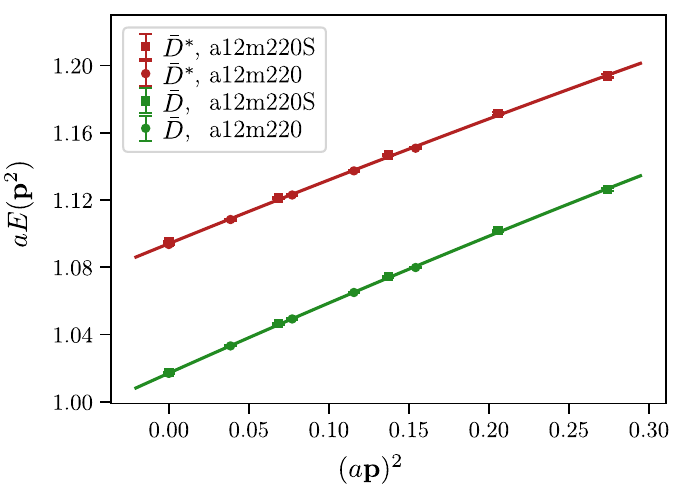}
 \caption{\label{fig:EDvspsqr}Results for $a E_{\bar{D}}(\mathbf{p}^2)$ and $a E_{\bar{D}^*}(\mathbf{p}^2)$ for $ 0 \leq \mathbf{p}^2 \leq 4 (2\pi/L)^2 $ from the two ensembles, along with fits of the form (\protect\ref{eq:dispRel_FERMI}). }
\end{figure}	

Our results for the $\bar{D}$ and $\bar{D}^*$ meson energies as a function of spatial momentum squared are shown in Fig.~\ref{fig:EDvspsqr}. We performed fits to the
combined data from the two ensembles using the three-parameter form
	\begin{equation}
		E_{\bar{D}^{(*)}}(\mathbf{p}^2)=E_{\bar{D}^{(*)}}(0)+\frac{\mathbf{p}^2}{2m_{\bar{D}^{(*)},{\rm kin}}}-\frac{\mathbf{p}^4}{8m_{\bar{D}^{(*)},4}^3}
		\label{eq:dispRel_FERMI}
	\end{equation}
to allow for different values of $m_{\bar{D}^{(*)},{\rm kin}}$, and $m_{\bar{D}^{(*)},4}$ due to discretization errors. Higher powers of $\mathbf{p}$ are expected to be negligible for the momentum range we use. The fit results are given in Table \ref{tab:singlemeson}. The spin-averaged kinetic mass agrees with the experimental value \cite{ParticleDataGroup:2022pth} within $3\%$, confirming the successful tuning of the charm-quark mass according to the Fermilab method \cite{El-Khadra:1996wdx}. We also find that the results for $m_{\bar{D}^{(*)},4}$ are actually consistent with $m_{\bar{D}^{(*)},{\rm kin}}$ within the statistical uncertainties.
	
	\begin{table*}
		\centering
		\begin{tabular}{cccccc}
			\hline\hline
			    $aE_{\bar{D}}(0)$    & $aE_{\bar{D}^*}(0) $  & $am_{\bar{D},\textrm{kin}}$ & $a m_{\bar{D}^*,\textrm{kin}} $ &  $ am_{\bar{D},4} $  & $ am_{\bar{D}^*,4} $ \\ \hline
			   $ 1.01718(38) $ & $ 1.09434(60) $ &     $ 1.172(14) $     &       $ 1.294(25) $       & $ 1.09(9) $ &  $ 1.18(16) $  \\ \hline\hline
		\end{tabular}
		\caption{\label{tab:singlemeson}$ \bar{D} $ and $ \bar{D}^* $ meson dispersion-relation parameters in lattice units, obtained from combined fits to the data from the a12m220S and a12m220 ensembles.}
	\end{table*}
	
For the $B$ and $B^*$ mesons, we did not expect a significant difference between $m_{B^{(*)},{\rm kin}}$, and $m_{B^{(*)},4}$ due to the high level of improvement of the lattice NRQCD action \cite{HPQCD:2011qwj}, and we therefore performed two-parameter fits of the form
\begin{equation}
  E_{B^{(*)}}(\mathbf{p}^2)=E_{B^{(*)}}(0)+\sqrt{m_{B^{(*)},{\rm kin}}^2+\mathbf{p}^2}-m_{B^{(*)},{\rm kin}}. \label{eq:Bdisprel}
\end{equation}

	\begin{table*}[h!]
		\centering
		\begin{tabular}{cccc}
			\hline\hline
			  $aE_B(0)$     & $aE_{B^*}(0) $    & $am_{B,\textrm{kin}}$ & $a m_{B^*,\textrm{kin}} $  \\ \hline
			  $ 0.4823(12) $               &    $ 0.5077(13) $               & $3.121(84)$ & $3.091(89)$ \\ \hline\hline
		\end{tabular}
		\caption{\label{tab:singlemesonbottom}$ B $ and $ B^* $ meson dispersion-relation parameters in lattice units, obtained from combined fits to the data from the a12m220S, a12m220, and a12m220L ensembles \cite{Alexandrou:2024iwi}. }
	\end{table*}

These fits included also a third ensemble of gauge configurations, a12m220L, with the same bare parameters and an even larger volume, and are discussed in more detail in Ref.~\cite{Alexandrou:2024iwi}. The data from all three ensembles are well-described jointly by Eq.~(\ref{eq:Bdisprel}) with the parameters given in Table \ref{tab:singlemesonbottom}. The spin-averaged kinetic $B$-meson mass is within 4\% of the experimental value \cite{ParticleDataGroup:2022pth}.
		
\subsection{\texorpdfstring{$\bm{\bar{b}\bar{c}ud}$}{bbar-cbar-u-d} energies and \texorpdfstring{$\bm{B^{(*)}}$-$\bm{\bar{D}}$}{B(*)-Dbar} scattering phase shifts}

\begin{figure*}
\includegraphics[width=0.4\linewidth]{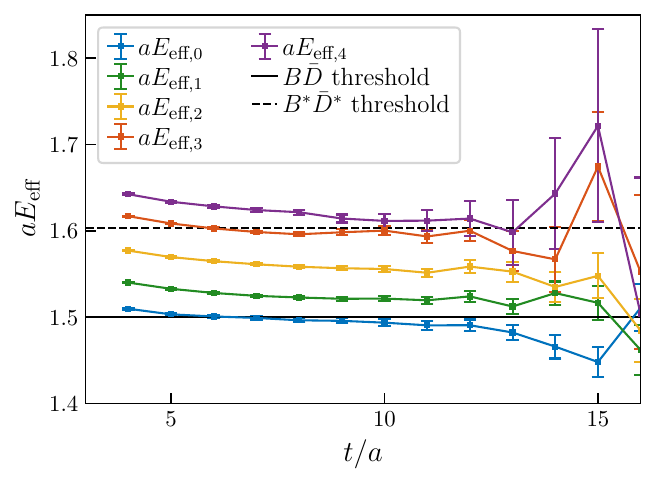} \includegraphics[width=0.4\linewidth]{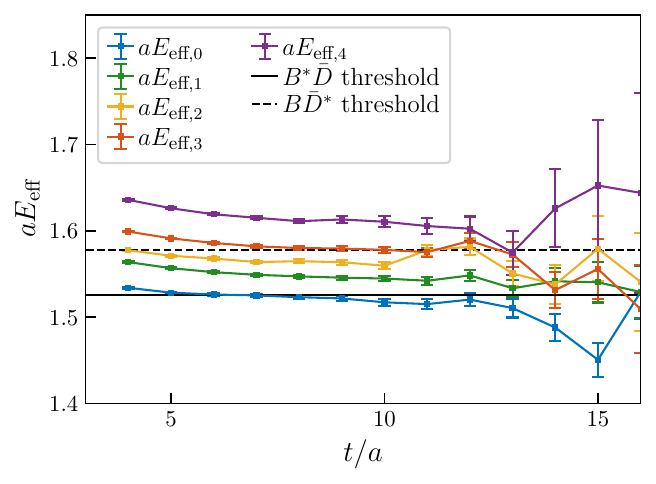} \\

\includegraphics[width=0.4\linewidth]{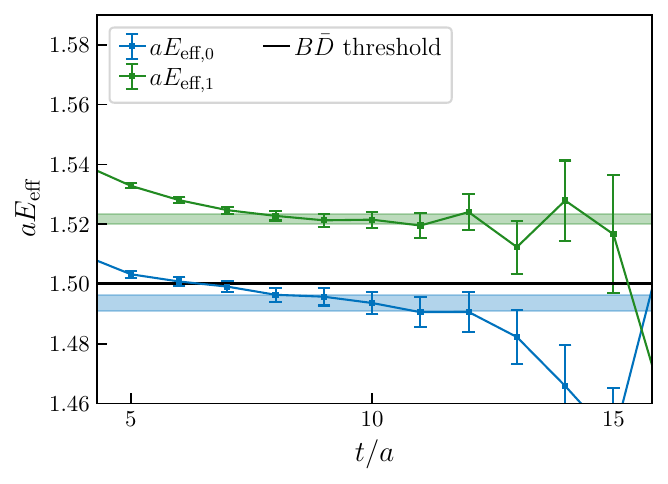} \includegraphics[width=0.4\linewidth]{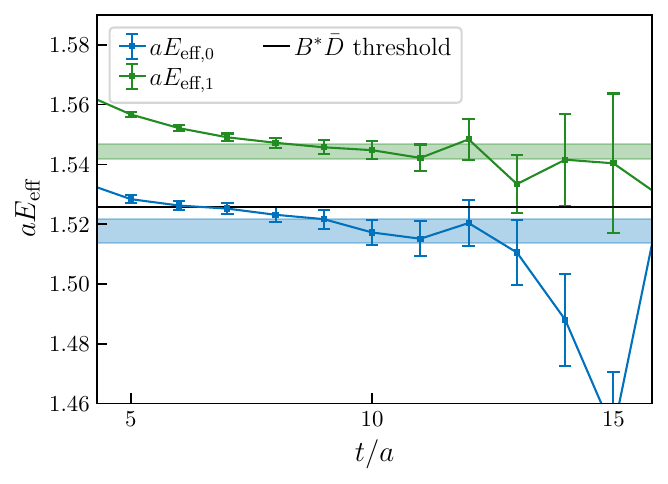} \\

\caption{\label{fig:Eeff}{\it Top left:} Effective energies $aE_{{\rm eff},n}(t)=\ln (\lambda_n(t,t_0)/\lambda_n(t+a,t_0))$ of the five lowest eigenvalues $\lambda_0(t,t_0),...,\lambda_4(t,t_0)$ of the $\bar{b}\bar{c}ud$ system with $I(J^P) = 0(0^+)$ on the a12m220S lattice, obtained by solving the GEVP for the $7\times7$ correlation matrix. For sufficiently large $t$, these effective energies correspond to the five lowest finite-volume energy levels with the given quantum numbers. Also shown are the strong-decay-threshold energies computed on the same lattice. Note that all absolute energies contain an overall constant shift due to the use of NRQCD and of the Fermilab method. {\it Top right:} The corresponding plot for the $\bar{b}\bar{c}ud$ system with $I(J^P) = 0(1^+)$, using the $8\times8$ correlation matrix. {\it Bottom left and right:} Close-up view of $aE_{{\rm eff},0}(t)$ and $aE_{{\rm eff},1}(t)$ together with the corresponding energy levels obtained by following the FLAG averaging procedure.}
\end{figure*}

Plots of the effective energies of the five lowest eigenvalues obtained from the GEVP for the $\bar{b}\bar{c}ud$ correlation matrices are shown in Fig.~\ref{fig:Eeff}. We determine the energy levels $E_n$ by carrying out correlated least-$\chi^2$ fits of the form $\lambda_n(t,t_0)=A_n e^{-E_n t}$. We perform fits for multiple different ranges $t_\text{min} \leq t \leq t_\text{max}$ with sufficiently large $t_\text{min}$ to ensure single-exponential behavior (we use all possible ranges with $8 \leq t_\text{min} / a \leq 10$, $15 \leq t_\text{max} / a \leq 19$ for $J = 0$, and $9 \leq t_\text{min} / a \leq 11$,  $15 \leq t_\text{max} / a \leq 19$ for $J = 1$; these ranges were chosen to yield $\chi^2/{\rm d.o.f.}$ in the vicinity of 1). Then we obtain the final estimate for $E_n$ from a weighted average that takes into account correlations and lowers the weights of fits with $\chi^2/{\rm d.o.f.}>1$, following the FLAG averaging procedure \cite{FlavourLatticeAveragingGroupFLAG:2021npn} (see also our discussion in Appendix B of Ref.~\cite{Meinel:2022lzo}). This weighting guarantees, in particular, that fits with too small $t_\text{min}$, for which excited states might not be negligible, are suppressed in the final result. The statistical uncertainties are calculated and propagated to the further analysis using jackknife.

Our results for the $\bar{b}\bar{c}ud$ finite-volume energies, their values relative to the threshold, the corresponding scattering momenta squared, and the corresponding products of scattering momentum and cotangent of $S$-wave scattering phase shifts are listed in Tables \ref{tab:bcudJ0} (for the $A_1^+$ irrep relevant for $J^P=0^+$) and \ref{tab:bcudJ1} (for the $T_1^+$ irrep relevant for $J^P=1^+$). As discussed in the main article, some high-lying energy levels are excluded from the phase-shift determination because they lie above inelastic thresholds, and the third energy level in the $T_1^+$ irrep is excluded because it corresponds to a state dominated by a $D$ wave; these cases are labeled ``N/A'' in the tables.

	\begin{table*}   
	 \begin{tabular}{ccccc}   
	  \hline\hline \\[-2.5ex]
	 Ensemble & $aE$ & $a\Delta E$ & $(ak)^2$ & $ak\cot\delta_0$ \\
	 \hline
	 a12m220S &  $ 1.4937(27) $    &     $-0.0065(23)$       &    $-0.0100(35)$   &   $-0.057(42)$   \\
	          &  $ 1.5217(16) $    &   $\wm0.0215(17)$       &  $\wm0.0378(24)$   & $\wm0.046(23)$   \\
	          &  $ 1.5568(17) $    &   $\wm0.0566(18)$       &  $\wm0.0987(26)$   &   $-0.001(42)$   \\
	          &  $ 1.5975(25) $    &   $\wm0.0973(28)$       &  $\wm0.1706(46)$   &   $-0.181(87)$   \\
	          &  $ 1.6172(36) $    &   $\wm0.1170(34)$       &       N/A   &        N/A          \\
\hline
	 a12m220  &  $ 1.4975(17) $    &     $-0.0014(17)$       &    $-0.0035(25)$   & $\wm0.008(90)$   \\
	          &  $ 1.5102(11) $    &   $\wm0.0112(13)$       &  $\wm0.0182(17)$   &   $-0.001(17)$   \\
	          &  $ 1.5318(11) $    &   $\wm0.0329(13)$       &  $\wm0.0554(17)$   &   $-0.004(35)$   \\
	          &  $ 1.5574(12) $    &   $\wm0.0585(14)$       &  $\wm0.0999(19)$   &   $-0.037(48)$   \\
	          &  $ 1.5750(29) $    &   $\wm0.0760(28)$       &  $\wm0.1307(43)$   &   $-0.13(11)\nb$    \\
	  \hline\hline
	 \end{tabular}
	 \caption{\label{tab:bcudJ0}The $\bar{b}\bar{c}ud$ finite-volume energies in the $A_1^+$ irrep, their values relative to the $B$-$\bar{D}$ threshold, the corresponding $B$-$\bar{D}$ scattering momenta squared, and the corresponding products of scattering momentum and cotangent of scattering phase shift (all in lattice units). Note that the absolute energies contain an overall constant offset due to the use of NRQCD and of the Fermilab method; this offset cancels in the differences to the threshold.}
	\end{table*}

	\begin{table*}   
	 \begin{tabular}{ccccc}   
	  \hline\hline \\[-2.5ex]
	 Ensemble & $aE$ & $a\Delta E$ & $(ak)^2$  & $ak\cot\delta_0$ \\
	 \hline
	 a12m220S &  $ 1.5176(40) $    &      $-0.0080(31)$      &     $-0.0122(49)$   &   $-0.080(44)$    \\
	          &  $ 1.5443(24) $    &    $\wm0.0187(22)$      &   $\wm0.0333(37)$   & $\wm0.007(29)$      \\
	          &  $ 1.5661(46) $    &    $\wm0.0404(31)$      &      N/A    &        N/A          \\
	          &  $ 1.5783(30) $    &    $\wm0.0526(26)$      &      N/A    &        N/A          \\
	          &  $ 1.6101(45) $    &    $\wm0.0844(41)$      &      N/A    &        N/A          \\
\hline
	 a12m220  &  $ 1.5217(19) $    &      $-0.0029(19)$      &     $-0.0053(28)$   &   $-0.037(48)$     \\
	          &  $ 1.5339(13) $    &    $\wm0.0093(14)$      &   $\wm0.0155(19)$   &   $-0.027(18)$     \\
	          &  $ 1.5456(16) $    &    $\wm0.0210(18)$      &      N/A    &        N/A          \\
	          &  $ 1.5553(13) $    &    $\wm0.0307(15)$      &   $\wm0.0522(19)$   &   $-0.074(43)$  \\
	          &  $ 1.5776(27) $    &    $\wm0.0530(26)$      &      N/A    &        N/A          \\
	  \hline\hline
	 \end{tabular}
	 	 \caption{\label{tab:bcudJ1}The $\bar{b}\bar{c}ud$ finite-volume energies in the $T_1^+$ irrep, their values relative to the $B^*$-$\bar{D}$ threshold, the corresponding $B^*$-$\bar{D}$ scattering momenta squared, and the corresponding products of scattering momentum and cotangent of scattering phase shift (all in lattice units). Note that the absolute energies contain an overall constant offset due to the use of NRQCD and of the Fermilab method; this offset cancels in the differences to the threshold.}
	\end{table*}
	
To demonstrate the importance of our operators and to verify the stability of our results, we also varied the set of operators entering the correlation matrix and the GEVP. For example, for ensemble a12m220 and $J = 0$, where we use the largest number of energy levels (the five lowest energy levels), we observed the following:
\begin{itemize}
\item It is important to include local operators: when all three local operators ($\op_1^{A_1^+}$, $\op_2^{A_1^+}$, $\op_3^{A_1^+}$) are omitted, the extracted second, third, and fourth energy levels are collectively shifted upward slightly, although the extracted ground-state energy remains essentially unchanged.

\item When only the diquark-antidiquark operator $\op_3^{A_1^+}$ is omitted, the extracted energy levels remain essentially unchanged.

\item When the scattering operator with the largest back-to-back momentum $\sqrt{3} \cdot 2 \pi / L$ is omitted, the extracted fourth energy level moves upward, to a position between the fourth and fifth energy levels obtained with the full operator set. This indicates that the number of scattering operators employed in this work is essential to study $k \cot\delta_0(k)$ up to the $B^\ast \bar{D}^\ast$ threshold, and raises some concern about the reliability of the extraction of the fifth energy level with the full operator basis. To test the stability of our determination of the ERE parameters (and hence of our predictions for the bound states and resonances), we performed an additional ERE fit for $J=0$ in which we omitted the fifth energy level from ensemble a12m220. This fit gives $a/a_0=-0.022(19)$, $r_0/(2a)=1.91(70)$, $b_0/a^3=-18.3(5.2)$, which agrees very well with the full fit (see Table \ref{tab:EREJ0params} below).
\end{itemize}

\FloatBarrier
\subsection{ERE fit parameters using the full momentum range}

This section contains our results for the ERE fit parameters for the main fits that use all available scattering momenta and phase shifts from Tables \ref{tab:bcudJ0} and \ref{tab:bcudJ1}. The fits were performed in lattice units,
	\begin{equation}
		ak \cot \delta_0 = \frac{a}{a_0} + \frac{r_0}{2a} (ak)^2 + \frac{b_0}{a^3} (ak)^4,
		\label{eq:Scatt_ERE_Quad_latunits}
	\end{equation}
where $a$ is the lattice spacing. We used the coefficients of $(ak)^0$, $(ak)^2$, and $(ak)^4$ as the fit parameters. Our results for these parameters, along with their correlation matrices, are given in Tables \ref{tab:EREJ0params} and \ref{tab:EREJ1params}. In addition, we provide the values of $1/a_0$, $r_0$, and $b_0$ in physical units in Table \ref{tab:EREparamsphys}.

In our ERE fits, we took into account the uncertainties of the lattice results for both $ak \cot \delta_0$ and $(ak)^2$ (the latter was done by promoting the $(ak)^2$ values to additional fit parameters). We also took into account the correlations between the different $ak \cot \delta_0$ values and the correlations between the different $(ak)^2$ values. However, we did not include the cross-correlations between $(ak)^2$ and  $ak \cot \delta_0$, which would lead to near-zero eigenvalues of the data covariance matrix whose inverse appears in the definition of $\chi^2$. To investigate how much this choice affects our results, we performed additional fits in which we also included those cross-correlations, but multiplied all off-diagonal elements of the data covariance matrix by a factor of 0.95 to avoid the issue with near-zero eigenvalues. These fits yield very similar results for both the central values and the uncertainties of the ERE parameters.

		\begin{table*}[h]   
	 \begin{tabular}{clccc}   
	  \hline\hline
	 Parameter & $\wm$Value &   \multicolumn{3}{c}{Correlation Matrix}   \\
	 \hline
	 $a/a_0$      & $-0.022(18)$   & $1$         & $-0.715$ & $\wm0.621$  \\
	 $r_0/(2a)$   & $\wm1.92(69)$  & $-0.715$    & $1$      & $-0.889$ \\
	 $b_0/a^3$    & $-18.4(5.0)$   & $\wm0.621$  & $-0.889$ & $1$      \\
	  \hline\hline
	 \end{tabular}
	 \caption{\label{tab:EREJ0params}Fit results for the $J=0$ ERE parameters and their correlation matrix.}
	\end{table*}

		\begin{table*}[h]   
	 \begin{tabular}{clccc}   
	  \hline\hline
	 Parameter & $\wm$Value &   \multicolumn{3}{c}{Correlation Matrix}   \\
	 \hline
	 $a/a_0$      &  $-0.044(20)$  & $1$         & $-0.725$ & $\wm0.594$  \\
	 $r_0/(2a)$   &  $\wm2.4(1.3)$  & $-0.725$  & $1$        & $-0.869$  \\
	 $b_0/a^3$    &  $-48(24)$  & $\wm0.594$ &  $-0.869$ & $1$      \\
	  \hline\hline
	 \end{tabular}
	 \caption{\label{tab:EREJ1params}Fit results for the $J=1$ ERE parameters and their correlation matrix.}
	\end{table*}

\begin{table*}[h]
	\centering
	\begin{tabular}{lcccc} \hline \hline
	&	$ 1/a_0 $ [fm$^{-1}$] & $ r_0 $ [fm] & $ b_0 $ [fm$^3$]  \\ \hline 
	$J\!=\!0$ &	$ -0.19(16) $ & $ 0.46(16) $ & $ -0.031(8)\phantom{0} $  \\ 
	$J\!=\!1$ &	$ -0.37(17) $ & $ 0.56(30) $ & $ -0.080(41) $  \\ \hline \hline
	\end{tabular}
	\caption{\label{tab:EREparamsphys}The ERE parameters in physical units.}
\end{table*}

\FloatBarrier
\subsection{ERE fits and bound-state poles using only the low-momentum region}

\begin{figure*}
\begin{center}
 \includegraphics[width=0.45\linewidth]{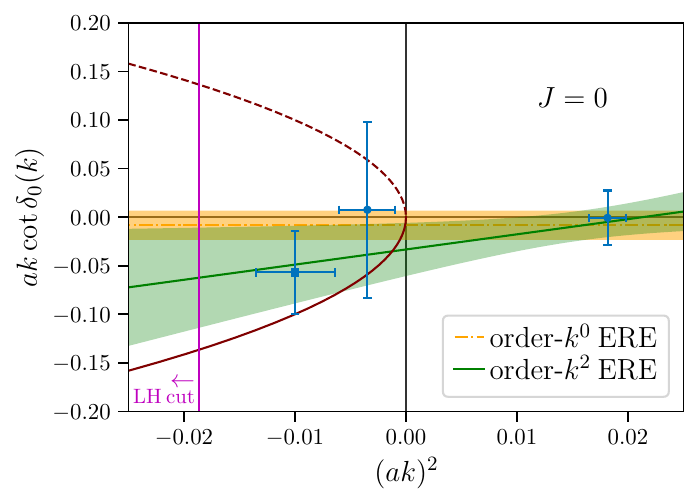} \hfill \includegraphics[width=0.45\linewidth]{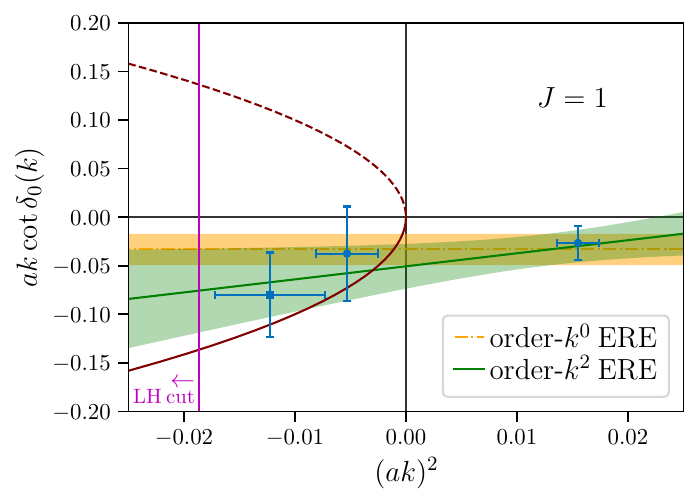}
 \end{center}
 \caption{\label{fig:ERE3points}ERE fits using only the three data points closest to the thresholds for $S$-wave $B$-$\bar{D}$ scattering (left) and $S$-wave $B^*$-$\bar{D}$ scattering (right); here, $a$ is the lattice spacing. Also shown are the functions $-\sqrt{-(ak)^2}$ (solid red parabolas) whose intersections with $ak \cot\delta_0(k)$ just below threshold correspond to the shallow $\bar{b}\bar{c}ud$ bound states we predict, and  $+\sqrt{-(ak)^2}$ (dashed red parabolas) whose intersections with $ak \cot\delta_0(k)$ would correspond to virtual $\bar{b}\bar{c}ud$ bound states. The vertical magenta lines show the positions of two-pion-exchange left-hand branch points.}
\end{figure*}

Our ERE fits to only the three data points closest to the thresholds are shown in Fig.~\ref{fig:ERE3points}. In this momentum region, 0th-order fits already describe the data well; for comparison, we also show fits through order $k^2$. The resulting ERE parameters, genuine-bound-state masses, and probabilities for the presence of a genuine bound state, a virtual bound state, or no bound state are given in Tables \ref{tab:EREparams3pointsorder0phys} and \ref{tab:EREparams3pointsorder1phys}.

\begin{table*}
	\centering
	\begin{tabular}{lcccccc} \hline \hline
	&	$ 1/a_0 $ [fm$^{-1}$] & $ \Delta m_{\rm GBS}$ [MeV] & GBS & VBS & NBS  \\ \hline 
	\vspace{-0.3cm} \\
	$J\!=\!0$ &  $-0.07(13)$   &  $-0.07_{-0.62}^{+0.05}$ & 71.9\% & 28.1\% & 0.0\%  \\ 
	\vspace{-0.3cm} \\
	$J\!=\!1$ &	 $-0.28(14)$   &  $-1.1_{-1.3}^{+0.7}$    & 98.2\% & 1.8\% & 0.0\%  \\
	\vspace{-0.3cm} \\ \hline \hline
	\end{tabular}
	\caption{\label{tab:EREparams3pointsorder0phys}Results for the inverse scattering lengths and genuine-bound-state masses (relative to the $B^{(*)}$-$\bar{D}$ thresholds) from the order-$k^0$ ERE fits to the three data points closest to the threshold.  Also shown are the fractions of the random ERE-parameter samples that yield a genuine bound state (GBS), a virtual bound state (VBS), or no bound state (NBS).}
\end{table*}

\begin{table*}
	\centering
	\begin{tabular}{lccccccc} \hline \hline
	&	$ 1/a_0 $ [fm$^{-1}$] & $ r_0 $ [fm] & $ \Delta m_{\rm GBS}$ [MeV]  & GBS & VBS & NBS \\ \hline 
	\vspace{-0.3cm} \\
	$J\!=\!0$ &	$-0.28(23)$  & $0.37(34)$ & $-1.2_{-3.0}^{+1.0}$  & 88.0\% & 12.0\% & 0.0\% \\ 
	\vspace{-0.3cm} \\
	$J\!=\!1$ & $-0.43(14)$	 & $0.32(30)$ & $-2.9_{-3.6}^{+2.1}$  & 98.4\% & 1.5\% & 0.2\% \\
  \vspace{-0.3cm} \\ \hline \hline
	\end{tabular}
	\caption{\label{tab:EREparams3pointsorder1phys}Results for the ERE parameters and genuine-bound-state masses (relative to the $B^{(*)}$-$\bar{D}$ thresholds) from the order-$k^2$ ERE fits to the three data points closest to the threshold. Also shown are the fractions of the random ERE-parameter samples that yield a genuine bound state (GBS), a virtual bound state (VBS), or no bound state (NBS). }
\end{table*}


\begin{thebibliography}{10}

\bibitem{Gell-Mann:1964ewy}
M.~Gell-Mann, ``{A Schematic Model of Baryons and Mesons},''
  \href{http://dx.doi.org/10.1016/S0031-9163(64)92001-3}{Phys. Lett. {\bfseries
  8} (1964) 214--215}.

\bibitem{Zweig:1964jf-xxx}
G.~Zweig, ``{An SU(3) model for strong interaction symmetry and its breaking.
  Version 2},'' in {\em Developments in the Quark Theory of Hadrons. Vol. 1.
  1964 - 1978}, D.~Lichtenberg and S.~P. Rosen, eds., pp.~22--101.
\newblock
1964.
\newblock
%%CITATION = CERN-TH-412;%%.

\bibitem{Jaffe:1976ig}
R.~L. Jaffe, ``{Multi-Quark Hadrons. 1. The Phenomenology of (2 Quark 2
  anti-Quark) Mesons},'' \href{http://dx.doi.org/10.1103/PhysRevD.15.267}{Phys.
  Rev. D {\bfseries 15} (1977) 267}.

\bibitem{Lebed:2016hpi}
R.~F. Lebed, R.~E. Mitchell, and E.~S. Swanson, ``{Heavy-Quark QCD Exotica},''
  \href{http://dx.doi.org/10.1016/j.ppnp.2016.11.003}{Prog. Part. Nucl. Phys.
  {\bfseries 93} (2017) 143--194},
  \href{http://arxiv.org/abs/1610.04528}{{\ttfamily arXiv:1610.04528
  [hep-ph]}}.

\bibitem{Chen:2022asf}
H.-X. Chen, W.~Chen, X.~Liu, Y.-R. Liu, and S.-L. Zhu, ``{An updated review of
  the new hadron states},''
  \href{http://dx.doi.org/10.1088/1361-6633/aca3b6}{Rept. Prog. Phys.
  {\bfseries 86} no.~2, (2023) 026201},
  \href{http://arxiv.org/abs/2204.02649}{{\ttfamily arXiv:2204.02649
  [hep-ph]}}.

\bibitem{LHCb:2021vvq}
{\bfseries LHCb} Collaboration, R.~Aaij {\em et~al.}, ``{Observation of an
  exotic narrow doubly charmed tetraquark},''
  \href{http://dx.doi.org/10.1038/s41567-022-01614-y}{Nature Phys. {\bfseries
  18} no.~7, (2022) 751--754},
  \href{http://arxiv.org/abs/2109.01038}{{\ttfamily arXiv:2109.01038
  [hep-ex]}}.

\bibitem{LHCb:2021auc}
{\bfseries LHCb} Collaboration, R.~Aaij {\em et~al.}, ``{Study of the doubly
  charmed tetraquark $T_{cc}^{+}$},''
  \href{http://dx.doi.org/10.1038/s41467-022-30206-w}{Nature Commun. {\bfseries
  13} no.~1, (2022) 3351}, \href{http://arxiv.org/abs/2109.01056}{{\ttfamily
  arXiv:2109.01056 [hep-ex]}}.

\bibitem{Ali:2018xfq}
A.~Ali, Q.~Qin, and W.~Wang, ``{Discovery potential of stable and
  near-threshold doubly heavy tetraquarks at the LHC},''
  \href{http://dx.doi.org/10.1016/j.physletb.2018.09.018}{Phys. Lett. B
  {\bfseries 785} (2018) 605--609},
  \href{http://arxiv.org/abs/1806.09288}{{\ttfamily arXiv:1806.09288
  [hep-ph]}}.

\bibitem{Francis:2016hui}
A.~Francis, R.~J. Hudspith, R.~Lewis, and K.~Maltman, ``{Lattice Prediction for
  Deeply Bound Doubly Heavy Tetraquarks},''
  \href{http://dx.doi.org/10.1103/PhysRevLett.118.142001}{Phys. Rev. Lett.
  {\bfseries 118} no.~14, (2017) 142001},
  \href{http://arxiv.org/abs/1607.05214}{{\ttfamily arXiv:1607.05214
  [hep-lat]}}.

\bibitem{Junnarkar:2018twb}
P.~Junnarkar, N.~Mathur, and M.~Padmanath, ``{Study of doubly heavy tetraquarks
  in Lattice QCD},'' \href{http://dx.doi.org/10.1103/PhysRevD.99.034507}{Phys.
  Rev. D {\bfseries 99} no.~3, (2019) 034507},
  \href{http://arxiv.org/abs/1810.12285}{{\ttfamily arXiv:1810.12285
  [hep-lat]}}.

\bibitem{Leskovec:2019ioa}
L.~Leskovec, S.~Meinel, M.~Pflaumer, and M.~Wagner, ``{Lattice QCD
  investigation of a doubly-bottom $\bar{b} \bar{b} u d$ tetraquark with
  quantum numbers $I(J^P) = 0(1^+)$},''
  \href{http://dx.doi.org/10.1103/PhysRevD.100.014503}{Phys. Rev. D {\bfseries
  100} no.~1, (2019) 014503}, \href{http://arxiv.org/abs/1904.04197}{{\ttfamily
  arXiv:1904.04197 [hep-lat]}}.

\bibitem{Mohanta:2020eed}
P.~Mohanta and S.~Basak, ``{Construction of $bb\bar{u}\bar{d}$ tetraquark
  states on lattice with NRQCD bottom and HISQ up and down quarks},''
  \href{http://dx.doi.org/10.1103/PhysRevD.102.094516}{Phys. Rev. D {\bfseries
  102} no.~9, (2020) 094516}, \href{http://arxiv.org/abs/2008.11146}{{\ttfamily
  arXiv:2008.11146 [hep-lat]}}.

\bibitem{Meinel:2022lzo}
S.~Meinel, M.~Pflaumer, and M.~Wagner, ``{Search for $\bar{b}\bar{b}us$ and
  $\bar{b}\bar{c}ud$ tetraquark bound states using lattice QCD},''
  \href{http://dx.doi.org/10.1103/PhysRevD.106.034507}{Phys. Rev. D {\bfseries
  106} no.~3, (2022) 034507}, \href{http://arxiv.org/abs/2205.13982}{{\ttfamily
  arXiv:2205.13982 [hep-lat]}}.

\bibitem{Hudspith:2023loy}
R.~J. Hudspith and D.~Mohler, ``{Exotic tetraquark states with two $b$ quarks
  and $J^P=0^+$ and $1^+$ $B_s$ states in a nonperturbatively tuned lattice
  NRQCD setup},'' \href{http://dx.doi.org/10.1103/PhysRevD.107.114510}{Phys.
  Rev. D {\bfseries 107} no.~11, (2023) 114510},
  \href{http://arxiv.org/abs/2303.17295}{{\ttfamily arXiv:2303.17295
  [hep-lat]}}.

\bibitem{Aoki:2023nzp}
T.~Aoki, S.~Aoki, and T.~Inoue, ``{Lattice study on a tetraquark state $T_{bb}$
  in the HAL QCD method},''
  \href{http://dx.doi.org/10.1103/PhysRevD.108.054502}{Phys. Rev. D {\bfseries
  108} no.~5, (2023) 054502}, \href{http://arxiv.org/abs/2306.03565}{{\ttfamily
  arXiv:2306.03565 [hep-lat]}}.

\bibitem{Xing:2018bqt}
Y.~Xing and R.~Zhu, ``{Weak Decays of Stable Doubly Heavy Tetraquark States},''
  \href{http://dx.doi.org/10.1103/PhysRevD.98.053005}{Phys. Rev. D {\bfseries
  98} no.~5, (2018) 053005}, \href{http://arxiv.org/abs/1806.01659}{{\ttfamily
  arXiv:1806.01659 [hep-ph]}}.

\bibitem{Hernandez:2019eox}
E.~Hern\'andez, J.~Vijande, A.~Valcarce, and J.-M. Richard, ``{Spectroscopy,
  lifetime and decay modes of the $T^-_{bb}$ tetraquark},''
  \href{http://dx.doi.org/10.1016/j.physletb.2019.135073}{Phys. Lett. B
  {\bfseries 800} (2020) 135073},
  \href{http://arxiv.org/abs/1910.13394}{{\ttfamily arXiv:1910.13394
  [hep-ph]}}.

\bibitem{Francis:2018jyb}
A.~Francis, R.~J. Hudspith, R.~Lewis, and K.~Maltman, ``{Evidence for
  charm-bottom tetraquarks and the mass dependence of heavy-light tetraquark
  states from lattice QCD},''
  \href{http://dx.doi.org/10.1103/PhysRevD.99.054505}{Phys. Rev. D {\bfseries
  99} no.~5, (2019) 054505}, \href{http://arxiv.org/abs/1810.10550}{{\ttfamily
  arXiv:1810.10550 [hep-lat]}}.

\bibitem{Hudspith:2020tdf}
R.~J. Hudspith, B.~Colquhoun, A.~Francis, R.~Lewis, and K.~Maltman, ``{A
  lattice investigation of exotic tetraquark channels},''
  \href{http://dx.doi.org/10.1103/PhysRevD.102.114506}{Phys. Rev. D {\bfseries
  102} (2020) 114506}, \href{http://arxiv.org/abs/2006.14294}{{\ttfamily
  arXiv:2006.14294 [hep-lat]}}.

\bibitem{Padmanath:2023rdu}
M.~Padmanath, A.~Radhakrishnan, and N.~Mathur, ``{Bound isoscalar axial-vector
  $bc\bar u\bar d$ tetraquark $T_{bc}$ in QCD},''
  \href{http://arxiv.org/abs/2307.14128}{{\ttfamily arXiv:2307.14128
  [hep-lat]}}.

\bibitem{Luscher:1986pf}
M.~Lüscher, ``{Volume Dependence of the Energy Spectrum in Massive Quantum
  Field Theories. 2. Scattering States},''
  \href{http://dx.doi.org/10.1007/BF01211097}{Commun. Math. Phys. {\bfseries
  105} (1986) 153--188}.

\bibitem{Luscher:1990ck}
M.~Lüscher and U.~Wolff, ``{How to Calculate the Elastic Scattering Matrix in
  Two-dimensional Quantum Field Theories by Numerical Simulation},''
  \href{http://dx.doi.org/10.1016/0550-3213(90)90540-T}{Nucl. Phys. B
  {\bfseries 339} (1990) 222--252}.

\bibitem{Luscher:1990ux}
M.~Lüscher, ``{Two particle states on a torus and their relation to the
  scattering matrix},''
  \href{http://dx.doi.org/10.1016/0550-3213(91)90366-6}{Nucl. Phys. B
  {\bfseries 354} (1991) 531--578}.

\bibitem{Luscher:1991cf}
M.~Lüscher, ``{Signatures of unstable particles in finite volume},''
  \href{http://dx.doi.org/10.1016/0550-3213(91)90584-K}{Nucl. Phys. B
  {\bfseries 364} (1991) 237--251}.

\bibitem{Lee:2009rt}
S.~H. Lee and S.~Yasui, ``{Stable multiquark states with heavy quarks in a
  diquark model},''
  \href{http://dx.doi.org/10.1140/epjc/s10052-009-1140-x}{Eur. Phys. J. C
  {\bfseries 64} (2009) 283--295},
  \href{http://arxiv.org/abs/0901.2977}{{\ttfamily arXiv:0901.2977 [hep-ph]}}.

\bibitem{Chen:2013aba}
W.~Chen, T.~G. Steele, and S.-L. Zhu, ``{Exotic open-flavor $bc\bar{q}\bar{q}$,
  $bc\bar{s}\bar{s}$ and $qc\bar{q}\bar{b}$, $sc\bar{s}\bar{b}$ tetraquark
  states},'' \href{http://dx.doi.org/10.1103/PhysRevD.89.054037}{Phys. Rev. D
  {\bfseries 89} no.~5, (2014) 054037},
  \href{http://arxiv.org/abs/1310.8337}{{\ttfamily arXiv:1310.8337 [hep-ph]}}.

\bibitem{Karliner:2017qjm}
M.~Karliner and J.~L. Rosner, ``{Discovery of doubly-charmed $\Xi_{cc}$ baryon
  implies a stable ($b b \bar{u} \bar{d}$) tetraquark},''
  \href{http://dx.doi.org/10.1103/PhysRevLett.119.202001}{Phys. Rev. Lett.
  {\bfseries 119} no.~20, (2017) 202001},
  \href{http://arxiv.org/abs/1707.07666}{{\ttfamily arXiv:1707.07666
  [hep-ph]}}.

\bibitem{Sakai:2017avl}
S.~Sakai, L.~Roca, and E.~Oset, ``{Charm-beauty meson bound states from
  $B(B^*)D(D^*)$ and $B(B^*)\bar D(\bar D^*)$ interaction},''
  \href{http://dx.doi.org/10.1103/PhysRevD.96.054023}{Phys. Rev. D {\bfseries
  96} no.~5, (2017) 054023}, \href{http://arxiv.org/abs/1704.02196}{{\ttfamily
  arXiv:1704.02196 [hep-ph]}}.

\bibitem{Deng:2018kly}
C.~Deng, H.~Chen, and J.~Ping, ``{Systematical investigation on the stability
  of doubly heavy tetraquark states},''
  \href{http://dx.doi.org/10.1140/epja/s10050-019-00012-y}{Eur. Phys. J. A
  {\bfseries 56} no.~1, (2020) 9},
  \href{http://arxiv.org/abs/1811.06462}{{\ttfamily arXiv:1811.06462
  [hep-ph]}}.

\bibitem{Agaev:2018khe}
S.~S. Agaev, K.~Azizi, B.~Barsbay, and H.~Sundu, ``{Weak decays of the
  axial-vector tetraquark $T_{bb;\bar{u} \bar{d}}^{-}$},''
  \href{http://dx.doi.org/10.1103/PhysRevD.99.033002}{Phys. Rev. D {\bfseries
  99} no.~3, (2019) 033002}, \href{http://arxiv.org/abs/1809.07791}{{\ttfamily
  arXiv:1809.07791 [hep-ph]}}.

\bibitem{Carames:2018tpe}
T.~F. Caram\'es, J.~Vijande, and A.~Valcarce, ``{Exotic $bc\bar q\bar q$
  four-quark states},''
  \href{http://dx.doi.org/10.1103/PhysRevD.99.014006}{Phys. Rev. D {\bfseries
  99} no.~1, (2019) 014006}, \href{http://arxiv.org/abs/1812.08991}{{\ttfamily
  arXiv:1812.08991 [hep-ph]}}.

\bibitem{Yang:2019itm}
G.~Yang, J.~Ping, and J.~Segovia, ``{Doubly-heavy tetraquarks},''
  \href{http://dx.doi.org/10.1103/PhysRevD.101.014001}{Phys. Rev. D {\bfseries
  101} no.~1, (2020) 014001}, \href{http://arxiv.org/abs/1911.00215}{{\ttfamily
  arXiv:1911.00215 [hep-ph]}}.

\bibitem{Tan:2020ldi}
Y.~Tan, W.~Lu, and J.~Ping, ``{Systematics of $QQ{\bar{q}}{\bar{q}}$ in a
  chiral constituent quark model},''
  \href{http://dx.doi.org/10.1140/epjp/s13360-020-00741-w}{Eur. Phys. J. Plus
  {\bfseries 135} no.~9, (2020) 716},
  \href{http://arxiv.org/abs/2004.02106}{{\ttfamily arXiv:2004.02106
  [hep-ph]}}.

\bibitem{Guo:2021yws}
T.~Guo, J.~Li, J.~Zhao, and L.~He, ``{Mass spectra of doubly heavy tetraquarks
  in an improved chromomagnetic interaction model},''
  \href{http://dx.doi.org/10.1103/PhysRevD.105.014021}{Phys. Rev. D {\bfseries
  105} no.~1, (2022) 014021}, \href{http://arxiv.org/abs/2108.10462}{{\ttfamily
  arXiv:2108.10462 [hep-ph]}}.

\bibitem{Richard:2022fdc}
J.-M. Richard, A.~Valcarce, and J.~Vijande, ``{Doubly-heavy tetraquark bound
  states and resonances},''
  \href{http://dx.doi.org/10.1016/j.nuclphysbps.2023.01.014}{Nucl. Part. Phys.
  Proc. {\bfseries 324-329} (2023) 64--67},
  \href{http://arxiv.org/abs/2209.07372}{{\ttfamily arXiv:2209.07372
  [hep-ph]}}.

\bibitem{Liu:2023vrk}
X.-Y. Liu, W.-X. Zhang, and D.~Jia, ``{Doubly heavy tetraquarks: Heavy quark
  bindings and chromomagnetically mixings},''
  \href{http://dx.doi.org/10.1103/PhysRevD.108.054019}{Phys. Rev. D {\bfseries
  108} no.~5, (2023) 054019}, \href{http://arxiv.org/abs/2303.03923}{{\ttfamily
  arXiv:2303.03923 [hep-ph]}}.

\bibitem{Ebert:2007rn}
D.~Ebert, R.~N. Faustov, V.~O. Galkin, and W.~Lucha, ``{Masses of tetraquarks
  with two heavy quarks in the relativistic quark model},''
  \href{http://dx.doi.org/10.1103/PhysRevD.76.114015}{Phys. Rev. D {\bfseries
  76} (2007) 114015}, \href{http://arxiv.org/abs/0706.3853}{{\ttfamily
  arXiv:0706.3853 [hep-ph]}}.

\bibitem{Eichten:2017ffp}
E.~J. Eichten and C.~Quigg, ``{Heavy-quark symmetry implies stable heavy
  tetraquark mesons $Q_iQ_j \bar q_k \bar q_l$},''
  \href{http://dx.doi.org/10.1103/PhysRevLett.119.202002}{Phys. Rev. Lett.
  {\bfseries 119} no.~20, (2017) 202002},
  \href{http://arxiv.org/abs/1707.09575}{{\ttfamily arXiv:1707.09575
  [hep-ph]}}.

\bibitem{Park:2018wjk}
W.~Park, S.~Noh, and S.~H. Lee, ``{Masses of the doubly heavy tetraquarks in a
  constituent quark model},''
  \href{http://dx.doi.org/10.1016/j.nuclphysa.2018.12.019}{Nucl. Phys. A
  {\bfseries 983} (2019) 1--19},
  \href{http://arxiv.org/abs/1809.05257}{{\ttfamily arXiv:1809.05257
  [nucl-th]}}.

\bibitem{Braaten:2020nwp}
E.~Braaten, L.-P. He, and A.~Mohapatra, ``{Masses of doubly heavy tetraquarks
  with error bars},''
  \href{http://dx.doi.org/10.1103/PhysRevD.103.016001}{Phys. Rev. D {\bfseries
  103} no.~1, (2021) 016001}, \href{http://arxiv.org/abs/2006.08650}{{\ttfamily
  arXiv:2006.08650 [hep-ph]}}.

\bibitem{Lu:2020rog}
Q.-F. L\"u, D.-Y. Chen, and Y.-B. Dong, ``{Masses of doubly heavy tetraquarks
  $T_{QQ^\prime}$ in a relativized quark model},''
  \href{http://dx.doi.org/10.1103/PhysRevD.102.034012}{Phys. Rev. D {\bfseries
  102} no.~3, (2020) 034012}, \href{http://arxiv.org/abs/2006.08087}{{\ttfamily
  arXiv:2006.08087 [hep-ph]}}.

\bibitem{Song:2023izj}
Y.~Song and D.~Jia, ``{Mass spectra of doubly heavy tetraquarks in
  diquark\ensuremath{-}antidiquark picture},''
  \href{http://dx.doi.org/10.1088/1572-9494/acc019}{Commun. Theor. Phys.
  {\bfseries 75} no.~5, (2023) 055201},
  \href{http://arxiv.org/abs/2301.00376}{{\ttfamily arXiv:2301.00376
  [hep-ph]}}.

\bibitem{Detmold:2019ghl}
{\bfseries USQCD} Collaboration, W.~Detmold, R.~G. Edwards, J.~J. Dudek,
  M.~Engelhardt, H.-W. Lin, S.~Meinel, K.~Orginos, and P.~Shanahan, ``{Hadrons
  and Nuclei},'' \href{http://dx.doi.org/10.1140/epja/i2019-12902-4}{Eur. Phys.
  J. A {\bfseries 55} no.~11, (2019) 193},
  \href{http://arxiv.org/abs/1904.09512}{{\ttfamily arXiv:1904.09512
  [hep-lat]}}.

\bibitem{Woss:2018irj}
A.~Woss, C.~E. Thomas, J.~J. Dudek, R.~G. Edwards, and D.~J. Wilson,
  ``{Dynamically-coupled partial-waves in $\rho\pi$ isospin-2 scattering from
  lattice QCD},'' \href{http://dx.doi.org/10.1007/JHEP07(2018)043}{JHEP
  {\bfseries 07} (2018) 043}, \href{http://arxiv.org/abs/1802.05580}{{\ttfamily
  arXiv:1802.05580 [hep-lat]}}.

\bibitem{Abdel-Rehim:2017dok}
A.~Abdel-Rehim, C.~Alexandrou, J.~Berlin, M.~Dalla~Brida, J.~Finkenrath, and
  M.~Wagner, ``{Investigating efficient methods for computing four-quark
  correlation functions},''
  \href{http://dx.doi.org/10.1016/j.cpc.2017.06.021}{Comput. Phys. Commun.
  {\bfseries 220} (2017) 97--121},
  \href{http://arxiv.org/abs/1701.07228}{{\ttfamily arXiv:1701.07228
  [hep-lat]}}.

\bibitem{El-Khadra:1996wdx}
A.~X. El-Khadra, A.~S. Kronfeld, and P.~B. Mackenzie, ``{Massive fermions in
  lattice gauge theory},''
  \href{http://dx.doi.org/10.1103/PhysRevD.55.3933}{Phys. Rev. D {\bfseries 55}
  (1997) 3933--3957}, \href{http://arxiv.org/abs/hep-lat/9604004}{{\ttfamily
  arXiv:hep-lat/9604004}}.

\bibitem{Lepage:1992tx}
G.~P. Lepage, L.~Magnea, C.~Nakhleh, U.~Magnea, and K.~Hornbostel, ``{Improved
  nonrelativistic QCD for heavy quark physics},''
  \href{http://dx.doi.org/10.1103/PhysRevD.46.4052}{Phys. Rev. D {\bfseries 46}
  (1992) 4052--4067}, \href{http://arxiv.org/abs/hep-lat/9205007}{{\ttfamily
  arXiv:hep-lat/9205007}}.

\bibitem{Blossier:2009kd}
B.~Blossier, M.~Della~Morte, G.~von Hippel, T.~Mendes, and R.~Sommer, ``{On the
  generalized eigenvalue method for energies and matrix elements in lattice
  field theory},'' \href{http://dx.doi.org/10.1088/1126-6708/2009/04/094}{JHEP
  {\bfseries 04} (2009) 094}, \href{http://arxiv.org/abs/0902.1265}{{\ttfamily
  arXiv:0902.1265 [hep-lat]}}.

\bibitem{Raposo:2023nex}
A.~B.~a. Raposo and M.~T. Hansen, ``{The L\"uscher scattering formalism on the
  t-channel cut},'' \href{http://dx.doi.org/10.22323/1.430.0051}{PoS {\bfseries
  LATTICE2022} (2023) 051}, \href{http://arxiv.org/abs/2301.03981}{{\ttfamily
  arXiv:2301.03981 [hep-lat]}}.

\bibitem{Padmanath:2022cvl}
M.~Padmanath and S.~Prelovsek, ``{Signature of a Doubly Charm Tetraquark Pole
  in $DD^*$ Scattering on the Lattice},''
  \href{http://dx.doi.org/10.1103/PhysRevLett.129.032002}{Phys. Rev. Lett.
  {\bfseries 129} no.~3, (2022) 032002},
  \href{http://arxiv.org/abs/2202.10110}{{\ttfamily arXiv:2202.10110
  [hep-lat]}}.

\bibitem{Iritani:2017rlk}
T.~Iritani, S.~Aoki, T.~Doi, T.~Hatsuda, Y.~Ikeda, T.~Inoue, N.~Ishii,
  H.~Nemura, and K.~Sasaki, ``{Are two nucleons bound in lattice QCD for heavy
  quark masses? Consistency check with L\"uscher\textquoteright{}s finite
  volume formula},'' \href{http://dx.doi.org/10.1103/PhysRevD.96.034521}{Phys.
  Rev. D {\bfseries 96} no.~3, (2017) 034521},
  \href{http://arxiv.org/abs/1703.07210}{{\ttfamily arXiv:1703.07210
  [hep-lat]}}.

\bibitem{Clark:2009wm}
{\bfseries QUDA} Collaboration, M.~A. Clark, R.~Babich, K.~Barros, R.~C.
  Brower, and C.~Rebbi, ``{Solving Lattice QCD systems of equations using mixed
  precision solvers on GPUs},''
  \href{http://dx.doi.org/10.1016/j.cpc.2010.05.002}{Comput. Phys. Commun.
  {\bfseries 181} (2010) 1517--1528},
  \href{http://arxiv.org/abs/0911.3191}{{\ttfamily arXiv:0911.3191 [hep-lat]}}.

\bibitem{Babich:2011np}
{\bfseries QUDA} Collaboration, R.~Babich, M.~A. Clark, B.~Joo, G.~Shi, R.~C.
  Brower, and S.~Gottlieb,
  \href{http://dx.doi.org/10.1145/2063384.2063478}{``{Scaling lattice QCD
  beyond 100 GPUs},''} in {\em {International Conference for High Performance
  Computing, Networking, Storage and Analysis}}.
\newblock 9, 2011.
\newblock \href{http://arxiv.org/abs/1109.2935}{{\ttfamily arXiv:1109.2935
  [hep-lat]}}.

\bibitem{Clark:2016rdz}
M.~A. Clark, B.~Jo\'o, A.~Strelchenko, M.~Cheng, A.~Gambhir, and R.~Brower,
  ``{Accelerating Lattice QCD Multigrid on GPUs Using Fine-Grained
  Parallelization},'' \href{http://arxiv.org/abs/1612.07873}{{\ttfamily
  arXiv:1612.07873 [hep-lat]}}.

\bibitem{Bhattacharya:2015wna}
{\bfseries PNDME} Collaboration, T.~Bhattacharya, V.~Cirigliano, S.~Cohen,
  R.~Gupta, A.~Joseph, H.-W. Lin, and B.~Yoon, ``{Iso-vector and Iso-scalar
  Tensor Charges of the Nucleon from Lattice QCD},''
  \href{http://dx.doi.org/10.1103/PhysRevD.92.094511}{Phys. Rev. D {\bfseries
  92} no.~9, (2015) 094511}, \href{http://arxiv.org/abs/1506.06411}{{\ttfamily
  arXiv:1506.06411 [hep-lat]}}.

\bibitem{Gupta:2018qil}
R.~Gupta, Y.-C. Jang, B.~Yoon, H.-W. Lin, V.~Cirigliano, and T.~Bhattacharya,
  ``{Isovector Charges of the Nucleon from 2+1+1-flavor Lattice QCD},''
  \href{http://dx.doi.org/10.1103/PhysRevD.98.034503}{Phys. Rev. D {\bfseries
  98} (2018) 034503}, \href{http://arxiv.org/abs/1806.09006}{{\ttfamily
  arXiv:1806.09006 [hep-lat]}}.

\bibitem{MILC:2012znn}
{\bfseries MILC} Collaboration, A.~Bazavov {\em et~al.}, ``{Lattice QCD
  Ensembles with Four Flavors of Highly Improved Staggered Quarks},''
  \href{http://dx.doi.org/10.1103/PhysRevD.87.054505}{Phys. Rev. D {\bfseries
  87} no.~5, (2013) 054505}, \href{http://arxiv.org/abs/1212.4768}{{\ttfamily
  arXiv:1212.4768 [hep-lat]}}.

\bibitem{ParticleDataGroup:2022pth}
{\bfseries Particle Data Group} Collaboration, R.~L. Workman {\em et~al.},
  ``{Review of Particle Physics},''
  \href{http://dx.doi.org/10.1093/ptep/ptac097}{PTEP {\bfseries 2022} (2022)
  083C01}.

\bibitem{HPQCD:2011qwj}
{\bfseries HPQCD} Collaboration, R.~J. Dowdall {\em et~al.}, ``{The Upsilon
  spectrum and the determination of the lattice spacing from lattice QCD
  including charm quarks in the sea},''
  \href{http://dx.doi.org/10.1103/PhysRevD.85.054509}{Phys. Rev. D {\bfseries
  85} (2012) 054509}, \href{http://arxiv.org/abs/1110.6887}{{\ttfamily
  arXiv:1110.6887 [hep-lat]}}.

\bibitem{Jansen:2008si}
{\bfseries ETM} Collaboration, K.~Jansen, C.~Michael, A.~Shindler, and
  M.~Wagner, ``{The Static-light meson spectrum from twisted mass lattice
  QCD},'' \href{http://dx.doi.org/10.1088/1126-6708/2008/12/058}{JHEP
  {\bfseries 12} (2008) 058}, \href{http://arxiv.org/abs/0810.1843}{{\ttfamily
  arXiv:0810.1843 [hep-lat]}}.

\bibitem{Alexandrou:2024iwi}
C.~Alexandrou, J.~Finkenrath, T.~Leontiou, S.~Meinel, M.~Pflaumer, and
  M.~Wagner, ``{$\bar b \bar b u d$ and $\bar b \bar b u s$ tetraquarks from
  lattice QCD using symmetric correlation matrices with both local and
  scattering interpolating operators},''
  \href{http://arxiv.org/abs/2404.03588}{{\ttfamily arXiv:2404.03588
  [hep-lat]}}.

\bibitem{FlavourLatticeAveragingGroupFLAG:2021npn}
{\bfseries Flavour Lattice Averaging Group (FLAG)} Collaboration, Y.~Aoki {\em
  et~al.}, ``{FLAG Review 2021},''
  \href{http://dx.doi.org/10.1140/epjc/s10052-022-10536-1}{Eur. Phys. J. C
  {\bfseries 82} no.~10, (2022) 869},
  \href{http://arxiv.org/abs/2111.09849}{{\ttfamily arXiv:2111.09849
  [hep-lat]}}.

\end{thebibliography}
\end{document}